\documentclass[12pt,preprint]{aastex}
\shorttitle{}
\shortauthors{Nesvorn\'y et al.}
\begin{document}
\title{STATISTICAL STUDY OF THE EARLY SOLAR SYSTEM'S INSTABILITY WITH 4, 5 AND 6 GIANT PLANETS}
\author{David Nesvorn\'y$^{1,2}$ and Alessandro Morbidelli$^2$}
\affil{(1) Department of Space Studies, Southwest Research Institute, Boulder, CO 80302, USA} 
\affil{(2) D\'epartement Cassiop\'ee, University of Nice, CNRS, Observatoire de la C\^ote d'Azur, Nice, 
06304, France}  

\begin{abstract}
Several properties of the Solar System, including the wide radial spacing and orbital eccentricities 
of giant planets, can be explained if the early Solar System evolved through a dynamical instability
followed by migration of planets in the planetesimal disk. Here we report the results of a statistical 
study, in which we performed nearly $10^4$ numerical simulations of planetary instability starting from 
hundreds of different initial conditions. We found that the dynamical evolution is typically too violent, 
if Jupiter and Saturn start in the 3:2 resonance, leading to ejection of at least one ice giant from the 
Solar System. Planet ejection can be avoided if the mass of the transplanetary disk of planetesimals was 
large ($M_{\rm disk}\gtrsim50$ M$_{\rm Earth}$), but we found that a massive disk would lead to 
excessive dynamical damping (e.g., final $e_{\rm 55} \lesssim0.01$ compared to present 
$e_{\rm 55}=0.044$, where $e_{55}$ is the amplitude of the fifth eccentric mode in the Jupiter's orbit), and 
to smooth migration that violates constraints from the survival of the terrestrial planets. 
Better results were obtained when the Solar System was assumed to have five giant planets initially and 
one ice giant, with the mass comparable to that of Uranus and Neptune, was ejected into interstellar space 
by Jupiter. The best results were obtained when the ejected planet was placed into the external 3:2 or 4:3 
resonance with Saturn and $M_{\rm disk} \simeq 20$ M$_{\rm Earth}$. The range of possible outcomes is rather 
broad in this case, indicating that the present Solar System is neither a typical nor expected result for 
a given initial state, and occurs, in best cases, with only a $\simeq$5\% probability (as defined by the 
success criteria described in the main text). The case with six giant planets shows interesting dynamics 
but does offer significant advantages relative to the five planet case.
\end{abstract}

\section{Introduction}

As giant planets radially migrate in the protoplanetary nebula they should commonly be drawn into compact systems, 
in which the pairs of neighbor planets are locked in the orbital resonances (Kley 2000, Masset \& Snellgrove 2001). 
The resonant planetary systems emerging from the protoplanetary disks can become dynamically unstable after the gas 
disappears, leading to a phase when planets scatter each other. This model can help to explain the observed resonant 
exoplanets (e.g., Gliese 876; Marcy et al. 2001), commonly large exoplanet eccentricities (Weidenschilling \& Marzari 
1996, Rasio \& Ford 1996), and microlensing data that show evidence for a large number of planets that are 
free-floating in interstellar space (Sumi et al. 2011; but see Veras \& Raymond 2012).

The Solar System, with the widely spaced and nearly circular orbits of the giant planets, bears little resemblance to 
the bulk of known exoplanets. Yet, if our understanding of physics of planet--gas-disk interaction is correct, it is 
likely that the young Solar System followed the evolutionary path outlined above. Jupiter and Saturn, for example, 
were most likely trapped in the 3:2 resonance (Masset \& Snellgrove 2001, Morbidelli \& Crida 2007, Pierens \& Nelson 
2008, Pierens \& Raymond 2011, Walsh et al. 2011), defined as $P_{\rm Sat}/P_{\rm Jup}=1.5$, where $P_{\rm Jup}$ and 
$P_{\rm Sat}$ are the orbital periods of Jupiter and Saturn (this ratio is 2.49 today). 

To stretch to its current configuration, the outer Solar System most likely underwent a dynamical instability during which 
Uranus and Neptune were scattered off of Jupiter and Saturn and acquired eccentric orbits (Thommes et al. 1999, Tsiganis 
et al. 2005, Morbidelli et al. 2007, Levison et al. 2011). The orbits were subsequently stabilized (and circularized) by 
damping the excess orbital energy into a massive disk of planetesimals located beyond the orbit of the outermost
planet (hereafter transplanetary disk), whose remains survived to this time 
in the Kuiper belt. Finally, as evidenced by dynamical structures observed in the Kuiper belt, planets radially migrated 
to their current orbits by scattering planetesimals (Malhotra et al. 1995, Gomes et al. 2004, Levison et al. 2008).

Several versions of the Solar System instability were proposed. Thommes et al. (1999), motivated by the apparent 
inability of the existing formation models to accrete Uranus and Neptune at their present locations, proposed that 
Uranus and Neptune
formed in the Jupiter-Saturn zone, and were scattered outwards when Jupiter, and perhaps Saturn, accreted nebular gas. 
This work represented an important paradigm shift in studies of the Solar System formation. Inspired by 
this work, Levison et al. (2001) suggested that the instability and subsequent dispersal of the planetesimal disk, if 
appropriately delayed (or due to the late formation of Uranus and Neptune; Wetherill 1975), could explain the Late Heavy 
Bombardment (LHB) of the Moon. The LHB was a spike in the bombardment rate some 4 Gyr ago suggested by the clustering
of ages of several lunar basins. The nature of the LHB, however, is still being debated (see Hartmann et al. 2000 and 
Chapman et al. 2007 for reviews).

In the {\it Nice} version of the instability (Tsiganis et al. 2005), Uranus and Neptune were assumed to have formed
just outside the orbit of Saturn, while Jupiter and Saturn were assumed to have initial orbits (where ``initial''
means at the time when the protoplanetary nebula dispersed), such that $P_{\rm Sat}/P_{\rm Jup}<2$. The instability was
triggered when Jupiter and Saturn, radially drifting by scattering planetesimals, approached and crossed the 2:1 
resonance ($P_{\rm Sat}/P_{\rm Jup}=2$). The subsequent evolution was similar to that found in the Thommes et al. 
model, but led to a better final orbital configuration of the planets (assuming that the disk contained
$\sim20$-50 M$_{\rm Earth}$ and was truncated at 30-35 AU). 

The {\it Nice} model is compelling because it explains the separations, eccentricities, and inclinations of the outer 
planets (Tsiganis et al. 2005), and many properties of the populations of small bodies
(Morbidelli et al. 2005, Nesvorn\'y et al. 2007, Levison et al. 2008, 2009, Nesvorn\'y \& Vokrouhlick\'y 2009).
It is currently the only migration model that is consistent with the current dynamical structure of the terrestrial planets 
(Brasser et al. 2009) and the main asteroid belt (Morbidelli et al. 2010). Moreover, the {\it Nice} model can 
also reproduce the magnitude and duration of the LHB (Gomes et al. 2005, Bottke et al. 2012). 

Two potentially problematic issues of the original {\it Nice} model (hereafter ONM) were addressed by Morbidelli et al. 
(2007) and Levison et al. (2011). Morbidelli et al. pointed out that the initial planetary orbits in the ONM were chosen
without the proper regard to the previous stage of evolution during which the giant planets interacted with the 
protoplanetary nebula. As shown by hydrodynamical studies (Masset \& Snellgrove 2001, Morbidelli \& Crida 2007, 
Pierens \& Nelson 2008, Pierens \& Raymond 2011), this interaction most likely led to the convergent migration of planets, and 
their trapping in orbital resonances. Morbidelli et al. (2007) studied the dynamical instability starting from the resonant 
planetary systems and showed that the orbit evolution of planets was similar to that of the ONM. 

To produce the LHB, the instability needs to occur late relative to the dispersal of the protoplanetary nebula. A 
protoplanetary nebula typically disperses in $\sim$3-10 Myr after the birth of the star (Haisch et al. 2001),
which for the Sun was probably contemporary to the formation of the first Solar System solids, 4.568 Gyr ago 
(Bouvier et al. 2007, Burkhardt et al. 2008). The onset of the LHB, traditionally considered to be 3.9-4.0 Gyr ago 
(Ryder 2002), has been recently revised to 4.1-4.2 Gyr ago (Bottke et al. 2012). This means that the instability had 
to occur with a delay of 350-650 Myr, with the exact value depending on the actual time of the LHB 
event.\footnote{Here we opted for including the full range of previous and new estimates of the delay, because it 
is not 100\% guaranteed that the new estimate is correct. Imposing the 350-450 My delay, instead of 350-650 My, 
would not make much of a difference, because most planetary systems that are stable over 450 My are also stable 
over 650 My.}
Such a late instability can be triggered in the ONM if the inner edge of 
the planetesimal disk was close, but not too close, to the outer ice giant. If the edge had been too close, the instability 
would have happened too early to be related to the LHB. If the edge had been too far, the instability would have happened 
too late or would not have happened at all, because the planetesimals would had stayed radially confined and not evolved 
onto planet-crossing orbits (where they could influence planets by short-range interactions). 

To avoid the need for fine tuning of the inner disk's edge in the ONM, Levison et al. (2011) proposed that the late 
instability was caused by the {\it long-range} interactions between planets and the radially-confined distant planetesimal 
disk. The long-term interactions arise when gravitational scattering between disk's planetesimals is taken into account.
As a result of these interactions, the planets and planetesimal disk slowly exchanged the energy and angular momentum 
until, after hundreds of Myr of small changes, the resonant locks between planets were broken and the instability
reigned supreme. Note that the instability trigger in Levison et al. (broken resonant locks) is fundamentally different 
from that of the ONM (major resonance crossing during migration).

Here we report the results of a new statistical study of the Solar System instability. Sections 2-4 explain our
method and constraints. The results are presented in Sect. 5. We discuss the plausible initial configurations of 
planets after the gas nebula dispersal, mass of the planetesimal disk, and effect of different trigger mechanisms on 
the results. The first steps in this direction were taken by Batygin \& Brown (2010), who found that some initial resonant 
configurations may work while others probably do not. 

Following Nesvorn\'y (2011) and Batygin et al. (2012) we also 
consider cases in which the young Solar System had extra ice giants initially and lost them during the scattering 
phase (these works and their relation to the present paper will be discussed in more detail below; see, e.g., Sect. 6).
We extend the previous studies of the Solar System instability by: (1) exploring a wide range of initial parameters 
(new resonant chains, six planets, etc.), (2) improving the statistics with up to 100 simulations per case, 
(3) considering different trigger mechanisms (including the one recently proposed by Levison et al. 2011), and 
(4) applying strict criteria of success/failure to all studied cases (see Sect. 3 and 4). Our conclusions are given 
in Sect. 7.
\section{Method}
We conducted computer simulations of the early evolution of the Solar System. In the first step (Phase 1), 
we performed hydrodynamic and 
$N$-body simulations to identify the resonant configurations that may have occurred among the young Solar System's giant 
planets. Our hydrodynamic simulations used {\it Fargo} (Masset 2000) and followed the method described in Morbidelli et al. 
(2007). As the {\it Fargo} simulations are CPU expensive, we used these results as a guide, and generated many additional 
resonant systems with the $N$-body integrator known as {\it SyMBA} (Duncan et al. 1998). 

Planets with masses corresponding to those of Jupiter, Saturn and ice giants, ordered in increasing orbital distance
from the Sun, were placed on initial orbits with the period ratios slightly larger that those of the selected resonances. 
We tested several cases for the initial radial order of ice giants including the case with Uranus on the inside of 
Neptune's orbit, Uranus on the outside of the Neptune's orbit, and case where both ice giants were given 
masses intermediate between that of Uranus and Neptune. The results (discussed in Sect. 5) obtained in these cases 
were statistically equivalent showing that the initial ordering of Uranus and Neptune was not important. 

The planets were then migrated into resonances with {\it SyMBA} modified to include forces that mimic the effects of gas. 
We considered cases with four, five and six initial planets, where the additional planets were placed onto resonant 
orbits between Saturn and the inner ice giant, or beyond the orbit of the outer ice giant. Additional planets were given 
the mass between 1/3 and 3 times the mass of Uranus.  

Different starting positions of planets, rates of the semimajor axis and eccentricity evolution (as implied by different
gas disk densities), and timescales for the gas disk's dispersal produced different results. For Jupiter and Saturn, we 
confined the scope of this study to the 3:2 and 2:1 resonances, because the former one is strongly preferred from previous 
hydrodynamic studies (Masset \& Snellgrove 2001, Morbidelli et al. 2007, Pierens \& Nelson 2008, Walsh et al. 2011, 
Pierens \& Raymond 2011). The 2:1 resonance was included for comparison. 
According to our tests with {\it Fargo}, trapping of Jupiter and Saturn in the 2:1 resonance may require special 
conditions (e.g., a low gas density implying slow convergent migration; Thommes et al. 2008). The 2:1 
resonance does not generically lead to the outward migration of Jupiter and Saturn that is required for the Grand 
Tack model (Walsh et al. 2011). 

The planetary eccentricities, inclinations and resonant amplitudes obtained at the end of our Phase-1 simulations were 
$e<0.1$, $i<0.2^\circ$ and $<60^\circ$. The inner ice giant typically had the most eccentric orbit ($0.05<e<0.1$), while 
the other planets' orbits were more circular (e.g., Jupiter had $e\lesssim0.01$ in most cases). We also considered cases 
with several different damping strengths for the same resonant chain. These cases helped us to understand the effect of
the initial excitation on the evolution of orbits during the instability. Changing the damping strength may not be strictly 
justified, because the semimajor axis migration and eccentricity damping are expected to be coupled (Goldreich \& Sari 
2003). The exact nature of this coupling, however, depends on several disk parameters that are poorly constrained (e.g., 
Lee \& Peale 2002). 

The instability of a planetary system can occur after the gas disk's dispersal when the stabilizing effects of gas are removed. 
Such an instability can be triggered spontaneously (e.g., Weidenschilling \& Marzari 1996), by divergent migration of 
planets produced by their interaction with planetesimals leaking into the planet-crossing orbits from a transplanetary 
disk (Tsiganis et al. 2005), or by long-range perturbations from a distant planetesimal disk (Levison et al. 2011).

It is often assumed that the instability in the Solar System occurred at the time of the LHB. As discussed in the introduction
this requires that the giant planets remained  on their initial resonant orbits for $\sim$350-650 Myr. 
To allow for this possibility, we examined the resonant configurations identified above and selected those that were 
stable over hundreds of Myr, if considered in isolation (i.e., no disk, planets only, no external perturbations). Only 
the stable systems were used for the follow-up simulations, in which we tracked the evolution of planetary orbits through 
and past the instability (Phase 2).

We included the effects of the transplanetary planetesimal disk in the Phase-2 simulations. The disk was represented by 1,000 
equal-mass bodies\footnote{Planets fully interact with each other and with the disk bodies. The gravitational interaction 
between disk bodies was neglected to cut down the CPU cost. This is justified because the behavior of the system 
{\it during} the instability is mainly driven by the interaction between planets, and between planets and planetesimals.} 
that were placed into orbits with low orbital eccentricities and inclinations (0.01), 
and radial distances between $r_{\rm in}<r<r_{\rm out}$. We also performed a limited number of simulations
with 100, 300, 3,000, and 10,000 disk bodies to test the effect of resolution,\footnote{Note that even our highest resolution 
runs did not have enough bodies to properly model the planetesimal disk, which is thought to have contained $\sim$1,000 
Pluto-size and myriads of smaller planetesimals (Morbidelli et al. 2009a). 
This should not be a major problem, however, because our results with increased
resolution showed only a minor dependence on the number of bodies used.} and with excited disks, where the  
eccentricities and inclinations of disk bodies were set to be $\simeq$0.1 (Sect. 5.2.1). The surface density was 
fixed at $\Sigma(r)=1/r$ in all simulations, because our tests showed that considering different density profiles has only 
a minimal effect on the results (see also Batygin \& Brown 2010). The outer edge of the disk was placed at $r_{\rm out}=30$ 
AU, so that the planetesimal-driven migration is expected to park Neptune near its present semimajor axis (Gomes et al. 
2004).\footnote{The real planetesimal disk probably continued all the way to $\simeq$47 AU, as implied by the existence
of the cold classical Kuiper belt (Batygin et al. 2011), but this extension likely contained too little mass to drive
Neptune's migration.}

We considered cases with $r_{\rm in}=a_n+\Delta$, where $a_n$ is the semimajor axis of the outer ice giant, $n$ is the number 
of planets, and $\Delta=0.5$-5 AU. The instability was usually triggered early for $\Delta < 1$ AU. To trigger the 
instability for $\Delta>1$ AU, we broke the resonant locks by altering the mean anomaly of one of the ice giants. This was 
done at the start of Phase-2 simulations. 
This method was inspired by the results of Levison et al. (2011). Different cases are denoted by ${\cal B}$ in the following 
text, where ${\cal B}(j)$ stands for breaking the resonant lock of ice giant planet $j$ (e.g., ${\cal B}(1)$ corresponds 
to the innermost ice giant; ${\cal B}(0)$ is the ONM trigger).   

In either case discussed above, the scattering phase between planets started shortly after the beginning of our simulations, 
which guaranteed low CPU cost. We considered different masses of the planetesimal disk, $M_{\rm disk}$, with $M_{\rm disk}$
between 10 and 100 M$_{\rm Earth}$. Thirty simulations were performed in each case, where different evolution histories were 
generated by randomly seeding the initial orbit distribution of planetesimals. The number of simulations was increased to 
100 in the interesting cases. In total, we completed nearly $10^4$ scattering simulations from over 200 different 
initial states. The new trials were selected with the knowledge of the results of the previous simulations and were 
optimized to sample the interesting parts of parameter space. Each system was followed for 100 Myr with the standard 
{\it SyMBA} integrator (Duncan et al. 1998), 
at which point the planetesimal disk was largely depleted and planetary migration ceased. We used a 
$h=0.5$ yr time step and verified that the results were statistically the same with $h=0.25$ yr.

\section{Constraints}

We defined four criteria to measure the overall success of simulations. First 
of all, the final planetary system must have four giant planets (criterion A) with orbits that resemble 
the present ones (criterion B). Note that A means that one and two planets must be ejected in the 
five- and six-planet planet cases, while all four planets need to survive in the four-planet case. 
As for B, we claim success if the final mean semimajor axis of each planet is within 20\% to its present value, 
and if the final mean eccentricities and mean inclinations are no larger than 0.11 and 2$^\circ$, 
respectively. These thresholds were obtained by doubling the current mean eccentricity of Saturn 
($e_{\rm Sat}=0.054$) and mean inclination of Uranus ($i_{\rm Ura}=1.02^\circ$; Table~1). 

For the successful runs, as defined above, we also checked on the history of encounters between giant planets, 
evolution of the secular $g_5$, $g_6$ and $s_6$ modes, and secular structure of the final planetary systems.
To explain the observed populations of the irregular moons that are roughly similar at each planet (when corrected 
for observational incompleteness; Jewitt \& Haghighipour 2007), all planets --including Jupiter-- must participate 
in encounters with other planets (Nesvorn\'y et al. 2007). 
Encounters of Jupiter and/or Saturn with ice giants may also be needed to excite the $g_5$ mode in the 
Jupiter's orbit to its current amplitude ($e_{55}=0.044$; Morbidelli et al. 2009b). Tables 2-4 list the secular 
frequencies and amplitudes of the outer Solar System planets.

It turns out that it is generally easy to have encounters of one of the ice giants to Jupiter if the planets
start in a compact resonant configuration (such encounters occur in most simulations for most cases tested here). 
The amplitudes of $g_6$ and $s_6$ modes also do not pose a problem. It is much harder to excite $e_{55}$, however. 
We therefore opt for a restrictive criterion in 
which we require that $e_{55}>0.022$ in the final systems, i.e., at least half of its current value (criterion C). The 
$e_{55}$ amplitude was determined by following the final planetary systems for additional 10 Myr (without
planetesimals), and Fourier-analyzing the results (\v{S}idlichovsk\'y \& Nesvorn\'y 
1996).

The evolution of secular modes, mainly $g_5$, $g_6$ and $s_6$, is constrained from their effects
on the terrestrial planets and asteroid belt. As giant planets scatter and migrate, these frequencies 
change. This may become a problem, if $g_5$ slowly swipes over the $g_1$ or $g_2$ modes, because the strong 
$g_1=g_5$ and $g_2=g_5$ resonances can produce excessive excitation and instabilities in the terrestrial 
planet system (Brasser et al. 2009, Agnor \& Lin 2011). The behavior of the $g_6$ and $s_6$ modes, on the 
other hand, is important for the asteroid belt (Morbidelli et al. 2010, Minton \& Malhotra 2011). 

As $g_5$, $g_6$ and $s_6$ are mainly a function of the orbital separation between Jupiter and Saturn, the 
constraints from the terrestrial planets and asteroid belt can be conveniently defined in terms of 
$P_{\rm Sat}/P_{\rm Jup}$. This ratio needs to evolve from $<$2.1 to $>$2.3 in $<1$ Myr (criterion D; also
final $P_{\rm Sat}/P_{\rm Jup}<2.8$; i.e., close to present $P_{\rm Sat}/P_{\rm Jup}=2.49$ but not too restrictive; 
see Sect. 4), which can be achieved, for example, if planetary encounters with 
an ice giant scatter Jupiter inward and Saturn outward. The 1-Myr limit is conservative in that a slower evolution 
of secular frequencies, which fails to satisfy the criterion D, would clearly violate the constraints. Note also 
that in most simulations that satisfy criterion D, $P_{\rm Sat}/P_{\rm Jup}$ evolves from $<$2.1 to $>$2.3 in 
much less than 1 Myr. Most simulations that satisfy D should therefore satisfy the constraints
from the terrestrial planets and asteroids.

In summary, our constraints A and B express the basic requirements on a successful model. Constraint C is 
more restrictive in that it places a more precise condition on the dynamical structure of the final systems.
Constraint D is the least rigorous. Given that it is related to the orbits of terrestrial planets, 
this constraint would not need to apply if the giant planet instability occurred before the terrestrial planets 
formed (i.e., early and well before the LHB). Note, however, that evolutions violating the constraint D 
would be difficult to reconcile with the present dynamical structure of the asteroid belt (Walsh \& Morbidelli 2011), 
regardless the timing of the giant planet instability. We therefore do not see a way around criterion D.

We report our results in Sect. 5. In doing so, we consider criteria C and D independently, but also discuss
the most interesting cases where both C and D were satisfied {\it simultaneously}. Note that the success rates
for C, D and C\&D were computed over the subset of simulations that satisfied A and B (i.e., it would 
not make sense, for example, to claim success for C if the system ended with incorrect number of planets). 
Also, B can be satisfied only if A is satisfied. The reported fractions were normalized by the total number 
of simulations performed in each case (equal to 30 or 100).

We ignored other constraints from the populations of small bodies in this work (e.g., cold classical Kuiper belt 
survival; Batygin et al. 2012). These constraints will be addressed in the upcoming work. 

\section{Measure of Success and Failure}

The instability is a chaotic process and as such it is not expected to produce deterministic results. 
Two planetary systems starting from the same initial conditions (that is, initial after 
the dispersal of the protoplanetary nebula) are expected to follow different evolutionary paths during the 
instability, and end up producing different final systems. It would thus be worthless to base our conclusions 
on one or a small number of simulations that may not be representative of the full range of possible 
outcomes. Here we performed 30 to 100 simulations for each initial setup. The number of simulations was 
chosen as a compromise between the need for reasonable statistics and our CPU limitations. 

A fundamental difficulty with studying the Solar System instability is that the Solar System, as we
know it now, is just {\it one} realization of {\it all} possible evolution paths. We do not know
if this realization is typical of those arising from the initial state, or the nature played a trick 
on us, and the Solar System has taken an unusual path. The latter possibility would offer a grim perspective 
on the possibility to determine the Solar System's state before the instability. Here we assume that the 
Solar System followed a path that was not ``exceptional''. Therefore we assume some cutoff probability 
for success in matching the criteria described in 
Sect. 3, and give preference to those initial conditions and evolution paths that exceed the cutoff. Now, 
the assumed cutoff must depend on how selective our criteria are because it is obvious that the measure of
the set of initial conditions that lead to the {\it exact} architecture of the Solar System is practically
zero. We therefore cannot set the criteria that would be too restrictive because we would not be able to
obtain satisfactory statistics with our 30-100 simulations per case. The criteria described in Sect. 3 were 
defined with this in mind. 

For example, there is nothing wrong with the initial setup that would lead to 3 or 4 giant planets in the end, 
as long as the two results occur with a roughly comparable likelihood. Statistical fluctuations are also expected 
for criteria B and C, because some systems may become more excited and/or radially spread (thus failing on B), 
while others end up dynamically cold (leaving Jupiter with low $e_{55}$ and failing C). Criterion D can be even 
more difficult to satisfy because planetary encounters produce a rather large variety of jumping Jupiter 
histories, in some of which Jupiter does not jump enough (leading to $P_{\rm Sat}/P_{\rm Jup}<2.3$) or jumps too 
much (leading to $P_{\rm Sat}/P_{\rm Jup} \gg 2.5$). Both these cases would fail D.

If we would optimistically assume that the success rate of 50\% for each of our four criteria would be 
satisfactory, the combined probability to simultaneously satisfy all of them would be $1/16 \simeq 6$\%. 
Based on this we set the target cutoff probability to be of the order of a few percent. The number of simulations
used in this work (see Sect. 2) was chosen so that we are able to measure such probabilities.

Note that in many cases studied here the differences between the model results and present Solar System are 
{\it systematic}. For example, in the runs starting with four planets, $e_{55}$ systematically ends up too low,
because the evolution needs to be relatively tranquil to satisfy A and B, and fails on C because $e_{55}$ is not 
excited enough (Sect. 5.1.1). It is clear that the probability to match our criteria is $\ll 1$\% in these 
cases (although we do not have enough statistics to say how low the success rate actually is).
\section{Results}
\subsection{Case with 4 Planets}
We start by discussing the results of simulations with four initial giant planets. All four planets have to survive 
in this case. We performed 2670 integrations for many different resonant chains, including cases of the 
2:1, 3:2, 4:3 and 5:4 resonances between pairs of planets. Table 5 summarizes the results of selected 
simulations. These cases were the most interesting ones, i.e. those where the criteria defined in Sect. 3 
were satisfied with the highest percentages (Tables 6 and 7 show the selected simulations for the 5- and 
6-planet cases). The second and higher order resonances were not considered. 
The results can be divided into two broad categories that share common traits: (1) Jupiter and Saturn starting 
in 3:2, and the inner ice giant in the 3:2 or higher degree resonance with Saturn (e.g., 4:3); and (2) Jupiter 
and Saturn in 3:2, and the inner ice giant in the 2:1 resonance with Saturn (or any other comparably distant orbit), 
or Jupiter and Saturn in the 2:1 resonance. We discuss the two categories in sections 5.1.1 and 5.1.2, respectively.
\subsubsection{3:2 and Tight}
In category (1), the system can easily be destabilized, and when it becomes unstable, the instability tends 
to be violent: one ice giant gets ejected from the Solar System or both ice giants survive, but their orbits end up 
being very excited ($e>0.1$) and/or scattered to large heliocentric distances ($a>50$ AU). This happens particularly
for disk masses $M_{\rm disk}\lesssim35$ M$_{\rm Earth}$. Thus the simulations with low disk masses are clearly unsuccessful 
in matching the present Solar System. 

For example, in the case with (3:2,3:2,4:3) very closely matching one of the systems discussed in Morbidelli 
et al. (2007), 87\% of simulations show ejection of one or more ice giants (for $M_{\rm disk}=35$ M$_{\rm Earth}$
and ${\cal B}(1)$). None of the remaining 13\% satisfied our criterion B, typically because 
the final orbits were overly excited (e.g., Uranus ended up with $i\sim10^\circ$). The simulations with 
$M_{\rm disk}<35$ M$_{\rm Earth}$ show even larger percentage of ejected planets and more excitation than the 
ones with  $M_{\rm disk}=35$ M$_{\rm Earth}$. Thus, disks with $M_{\rm disk}\lesssim 35$ M$_{\rm Earth}$ do not
apply. This result is robust in that it does not depend on the initial resonant chain (within the definition 
of category (1)).

\begin{figure}
\epsscale{1.0}
\plotone{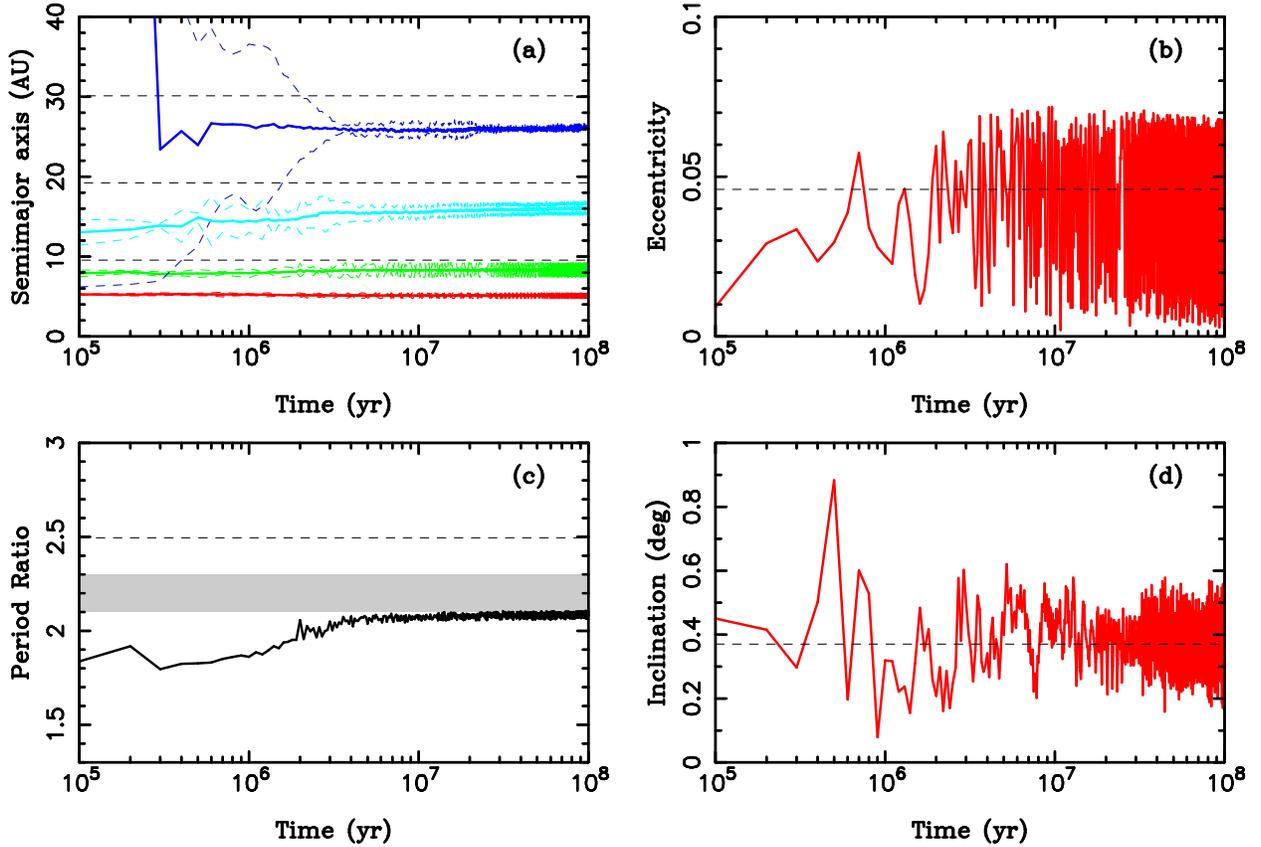}
\caption{Orbit histories of the giant planets in a simulation with four initial planets. The four planets were 
started in the (3:2,4:3,3:2) resonant chain, $M_{\rm disk}=35$ M$_{\rm Earth}$ and ${\cal B}(1)$ (see Sect. 2
for the definition of ${\cal B}(1)$). (a) The semimajor 
axes (solid lines), and perihelion and aphelion distances (dashed lines) of each planet's orbit. The red, green, 
turquoise and blue lines correspond to Jupiter, Saturn, Uranus and Neptune. The black dashed lines show
the semimajor axes of planets in the present Solar System. (c) The period ratio $P_{\rm Sat}/P_{\rm Jup}$. The dashed line 
shows $P_{\rm Sat}/P_{\rm Jup}=2.49$, corresponding to the period ratio in the present Solar System. The shaded area 
approximately denotes the zone where the secular resonances with the terrestrial planets occur. (b) Jupiter's 
eccentricity. (d) Jupiter's inclination. The dashed lines in (b) and (d) show the present mean eccentricity and 
inclination of Jupiter.}
%job14 in p4/r32c4/m2d1
\label{case1}
\end{figure}

A single exception from the above rule was one simulation started with (3:2,4:3,3:2), which was, perhaps
incidentally, a resonant chain favored by Batygin \& Brown (2010). In this particular run, Neptune was scattered
to $a\simeq26$ AU, where it was quickly circularized by dynamical friction from the planetesimal disk (Fig. \ref{case1}). 
The subsequent migration of Neptune and Uranus was minimal and the two orbits ended too close to the Sun
($a=16$ and 26 AU, respectively). 

On the positive side, Jupiter's orbit was excited by the encounters with Uranus, such that $\bar{e}_{\rm Jup} =0.042$ and 
$\bar{i}_{\rm Jup}=0.37^\circ$ in the end (bar indicates mean values; initial values were 0.002 and 0.013$^\circ$, respectively), 
both in a good agreement with the present values ($\bar{e}_{\rm Jup}=0.046$ and $\bar{i}_{\rm Jup}=0.37^\circ$; Table 1). Even 
more encouragingly, final $e_{55}=0.036$, also in a close agreement with present $e_{55}=0.044$ (Table 3). 
The main problem with this run, however, was that the jump in Jupiter and Saturn orbits produced by scattering 
encounters with Uranus and Neptune was minimal ($P_{\rm Sat}/P_{\rm Jup}$ moved to $\simeq1.8$ at $t<10^5$ yr) 
so that the final $P_{\rm Sat}/P_{\rm Jup}$ ratio was too low ($\simeq$2.1). 

It is difficult to evaluate whether this is a fundamental problem, or whether we would find better cases if we 
increased the number of simulations. We increased the number of simulations to 100 but did not obtain a case 
that would be even remotely as good as the one shown in Fig. \ref{case1}. We were therefore probably just lucky 
when finding this case with fewer runs. We conclude that it might be possible to match the A, B and C 
constraints with $M_{\rm disk}\approx35$~M$_{\rm Earth}$, but the likelihood of this occurring is $\lesssim$1\%.
If, in addition, criterion D is considered, the probability drops to $\ll$1\%. 

A better success in matching criteria A and B was obtained with more massive disks ($M_{\rm disk}\gtrsim 50$ 
M$_{\rm Earth}$). The success for A and B increased with $M_{\rm disk}$, because the more massive disks were capable 
of stabilizing the system of four planets more efficiently. The best success rate for A and B was obtained with 
$M_{\rm disk} = 75$ M$_{\rm Earth}$ and $r_{\rm out}=26$ AU, where 67\% of jobs matched A and 40\% satisfied B.
We set the outer disk edge at 26~AU, because in our initial runs with $r_{\rm out}=30$ AU Neptune migrated to 
$\sim$35 AU. A massive planetesimal disk is apparently capable of migrating Neptune beyond the initial outer 
disk's edge.

\begin{figure}
\epsscale{0.6}
\plotone{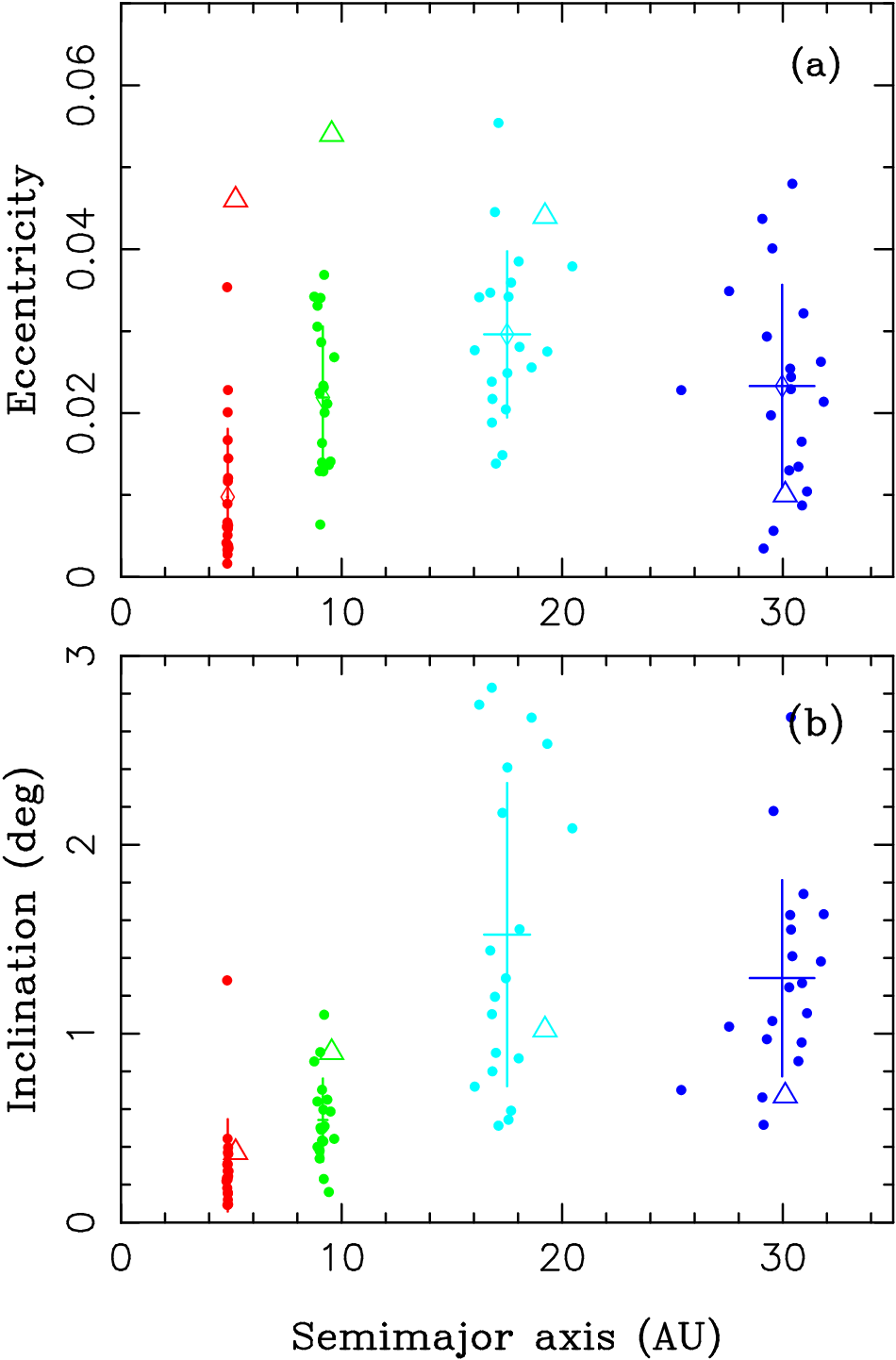}
\caption{Final orbits obtained in our simulations with four planets started in the (3:2,3:2,4:3) resonant 
chain, $M_{\rm disk}=75$ M$_{\rm Earth}$ and ${\cal B}(1)$ (see Sect. 2 for the definition of ${\cal B}(1)$). 
(a) Mean eccentricity. (b) Mean inclination. The mean
orbital elements were obtained by averaging the osculating orbital elements over the last 10 Myr (i.e., from 90 to 
100 Myr). Only the systems ending with four planets are plotted here (dots). The bars show the mean and standard deviation of 
the model distribution of orbital elements. The mean orbits of real planets are shown by triangles. Colors red, green, 
turquoise and blue correspond to Jupiter, Saturn, Uranus and Neptune.}
% job100/p4r32c1/m5d1/
\label{sum1}
\end{figure}

The overall distribution of final orbits was reasonable in this case (Fig. \ref{sum1}). Uranus ended up slightly inside its 
present orbit ($a_{\rm Ura}=17.5$ AU compared to real $a_{\rm Ura}=19.2$ AU), which may or may not be a problem.
On one hand, Uranus ended up too close to the Sun in most of the simulations that we performed with 4 planets and the 3:2 
resonance between Jupiter and Saturn. So, this is a systematic effect. On the other hand, one might be able 
to resolve this problem by modifying the initial setup (e.g., starting Uranus on a larger orbit; but see Sect. 
5.1.2). Alternatively, the present orbit of Uranus is an outlier in the semimajor axis distribution about 
2$\sigma$ from the mean. 

\begin{figure}
\epsscale{1.0}
\plotone{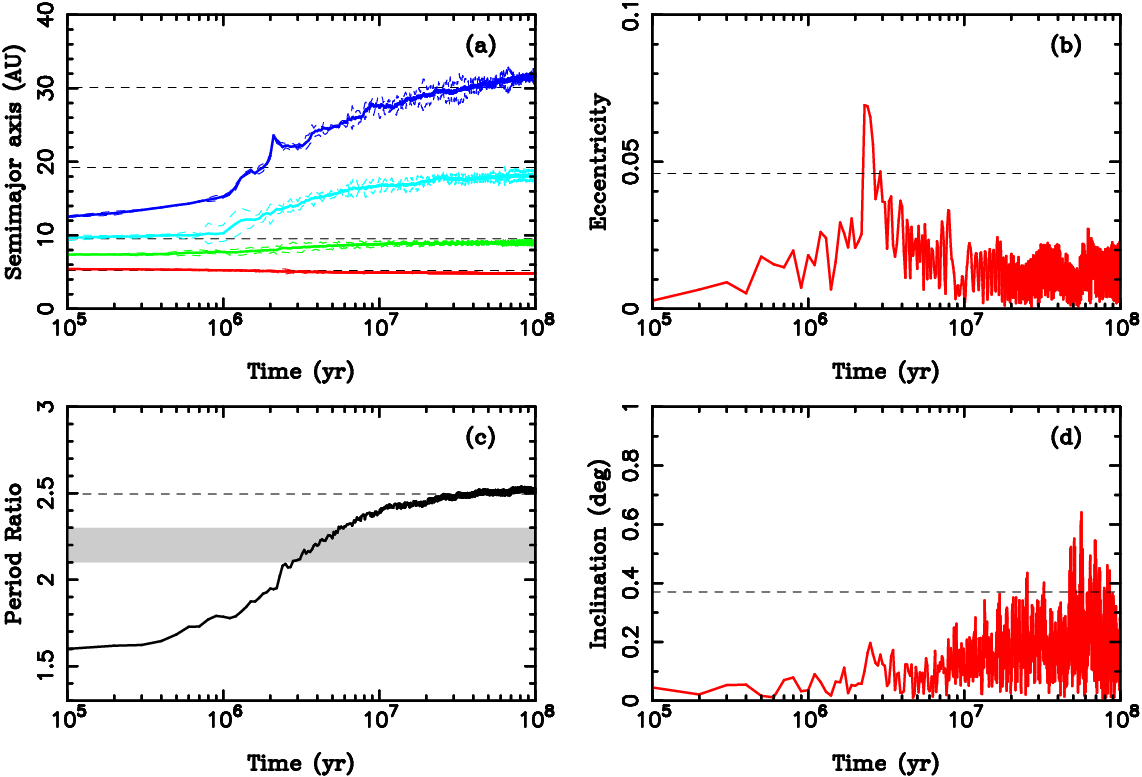}
\caption{Orbit histories of the giant planets in a simulation with four initial planets.
See the caption of Fig. \ref{case1} for the description of orbital parameters shown here. 
The four planets were started in the (3:2,3:2,4:3) resonant chain, and $M_{\rm disk}=75$ M$_{\rm Earth}$.}
%job number 10 from job100/p4r32c1
\label{case2}
\end{figure}

A more fundamental (and also systematic) problem of these simulations was that both Jupiter and Saturn ended up on 
orbits that were too circular. The mean eccentricities that we obtained in the simulations with $M_{\rm disk} = 75$ 
M$_{\rm Earth}$ were $\bar{e}_{\rm Jup}=0.01$ and $\bar{e}_{\rm Sat}=0.02$, while real $\bar{e}_{\rm Jup} = 0.046$ and 
$\bar{e}_{\rm Sat}=0.054$. This was a consequence of the damping effect of massive planetesimal disk on the gas giant 
orbits (see Fig. \ref{case2} for an illustration). This problem was general for all simulations that we performed 
with massive disks.

To understand this effect better, we analyzed the simulations with massive planetesimal disks in more detail.
We found that it is generally not a problem to excite $e_{\rm Jup}$ and $e_{\rm Sat}$ to $\sim$0.05 by scattering 
encounters with one of the ice giants. What is problematic, however, is to maintain the excited state when 
the massive planetesimal disk is present. In fact, in all our simulations with massive disks, $e_{\rm Jup}$ and 
$e_{\rm Sat}$ were quickly damped after the excitation event. 

We believe that the damping effect arises from {\it secular friction} (Levison et al. 2011) and occurs when $e$ and/or 
the longitude of perihelion 
$\varpi$ of a planet suddenly changes (e.g., by being scattered off of another planet or due to orbital resonance crossing). 
This changes the phase and amplitude of the secular forcing between the planet and the planetesimal disk. When 
the system relaxes the mean eccentricity of the planetesimals increases, while that of the planet decreases.  
The secular friction damps planet's inclination as well.\footnote{Secular friction is different from an effect
known as the dynamical friction (Binney \& Tremaine 1987), which arises from encounters between bodies.}

The excessive damping of $e_{\rm Jup}$ and $e_{\rm Sat}$ could potentially be avoided if the planetesimal disk were dynamically 
excited or depleted prior to the event that excites $e_{55}$. This might happen, for example, if Neptune were scattered deep into 
the planetesimal disk (to 28-30 AU) and disrupted it before the $e_{55}$ excitation event. Unfortunately, we did not see 
this happening in our simulations, except for a few cases that were flawed for other reasons (e.g., Fig. \ref{case1}). 
Another possibility, discussed in more detail in Sect. 5.2.1, is that the planetesimal disk was stirred by large objects 
that formed in the disk and/or by resonances with ice giants. Whatever the stirring mechanism was, however, we found that 
a disk that did not exert a strong damping effect on Jupiter and Saturn was not capable of preventing ejection of Uranus or 
Neptune. These simulations thus failed on A or C.

There were two additional problems with the simulations such as the one illustrated in Fig. \ref{case2}. First,
the ones that matched criteria A and B were typically those that avoided any strong interactions between planets. 
This resulted in a nearly smooth migration histories of planets that violated D. In the specific (and representative) case 
showed in Fig. \ref{case2}, the system spent more than 3 Myr with $2.1<P_{\rm Sat}/P_{\rm Jup}<2.3$, which is expected to
be incompatible with the survival of the terrestrial planets (as discussed in Sect. 3).

Second, the smooth crossing of the 2:1 resonance between Jupiter and Saturn excited $e_{56}$ but not 
$e_{55}$. The simulations therefore failed on C. The excitation by resonant crossing mainly occurs as the 
eccentricity of Saturn evolves over the resonance with Jupiter (Jupiter's eccentricity changes less because 
Jupiter has a larger mass). Thus, the resonant crossing directly affects the amplitudes related to the $g_6$ mode,
and not those of the $g_5$ mode (Morbidelli et al. 2009b).  The needed excitation of $e_{55}$ could more easily
be achieved by scattering encounters of an ice giant with Jupiter, but such encounters generally lead to
violent orbital histories that fail A or B if only four giant planets are considered.

\begin{figure}
\epsscale{1.0}
\plotone{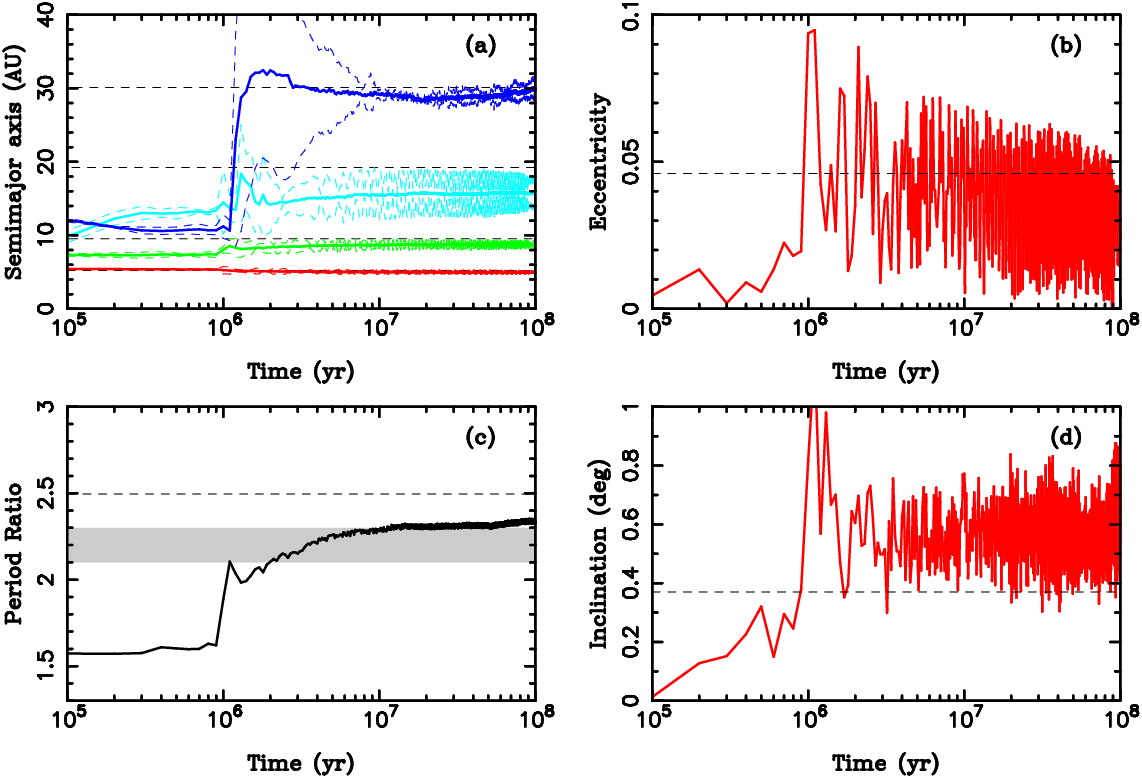}
\caption{Orbit histories of the giant planets in a simulation with four initial planets.
See the caption of Fig. \ref{case1} for the description of orbital parameters shown here. 
The four planets were started in the (3:2,3:2,4:3) resonant chain, $M_{\rm disk}=50$ 
M$_{\rm Earth}$ and $\Delta=1$ AU.} 
%(job number 8 from job100/p4r32c2)}
\label{case3}
\end{figure}

Finally, we discuss the most promising four-planet case obtained with (3:2,3:2,4:3) and $M_{\rm disk}=50$ 
M$_{\rm Earth}$. In this case, 27\% of runs satisfied A, 11\% of runs satisfied B, and we also found one case 
(out of a hundred) that satisfied criterion C (Fig. \ref{case3}). This specific simulation ended with $g_5=5.98$ 
arcsec yr$^{-1}$ and $g_6=32.4$ arcsec yr$^{-1}$, reflecting the fact that Jupiter and Saturn finished a bit closer to 
each other than in reality, and $e_{55}=0.026$,  $e_{56}=0.010$, $e_{66}=0.031$, and $e_{65}=0.022$. This is a 
pretty good match to the real values (Table 3), although $e_{55}$ is only a bit larger than a half of the present 
value. Unfortunately, although this simulations also showed additional encouraging signs (e.g., a $\sim$0.5 jump 
of $P_{\rm Sat}/P_{\rm Jup}$), it violated D, because the jump of $P_{\rm Sat}/P_{\rm Jup}$ was not large enough.

In summary, we were not able to find any initial condition for the four-planet case, where the likelihood of 
simultaneously matching all our four criteria would exceed 1\%.  
\subsubsection{3:2 and Loose, or 2:1}
Here we first consider cases where Jupiter and Saturn are in the 3:2 resonance, the inner ice giant in 
the 2:1 resonance with Saturn, and various resonances between the first and second ice giants. These cases showed a different 
behavior than the ones discussed in the previous section. They were more stable in that breaking the 
resonant locks of the inner ice giant did not always generate an instability. If it did, the instability
was usually mild such that close encounters between planets did not occur, and planets ended up migrating smoothly. 
This had two conflicting consequences. The smooth migration gave a large success rate for A and B, because the system 
did not suffer destabilizing perturbations and evolved quietly. In most other aspects, however, these simulations 
produced incorrect results.
   
First of all, Jupiter and Saturn did not reach the current $P_{\rm Sat}/P_{\rm Jup}=2.5$ unless the disk was heavy. 
This was because, in absence of encounters with ice giants, the gas giants did not jump, and all the 
migration work had to be accomplished by planetesimals. With heavy disks, which gave a more 
reasonable final $P_{\rm Sat}/P_{\rm Jup}$ ratio, $P_{\rm Sat}/P_{\rm Jup}$ slowly migrated over
2.1-2.3 and violated our constraint D. 

The eccentricities of Jupiter and Saturn were excited when these planets crossed their mutual 2:1 resonance. The
resonant crossing led to $e_{\rm Jup}\simeq 0.04$ and $e_{\rm Sat}\simeq 0.08$, which was lovely, except that 
this eccentricity excitation was contained in the secular modes related to $g_6$. The simulations therefore 
violated C as $e_{55}$ was never large enough. We have not found a single case where the constraint C 
{\it or} D were satisfied. We conclude that the resonant chains with the inner ice giant in 
the 2:1 resonance with Saturn can be ruled out.

\begin{figure}
\epsscale{0.6}
 \plotone{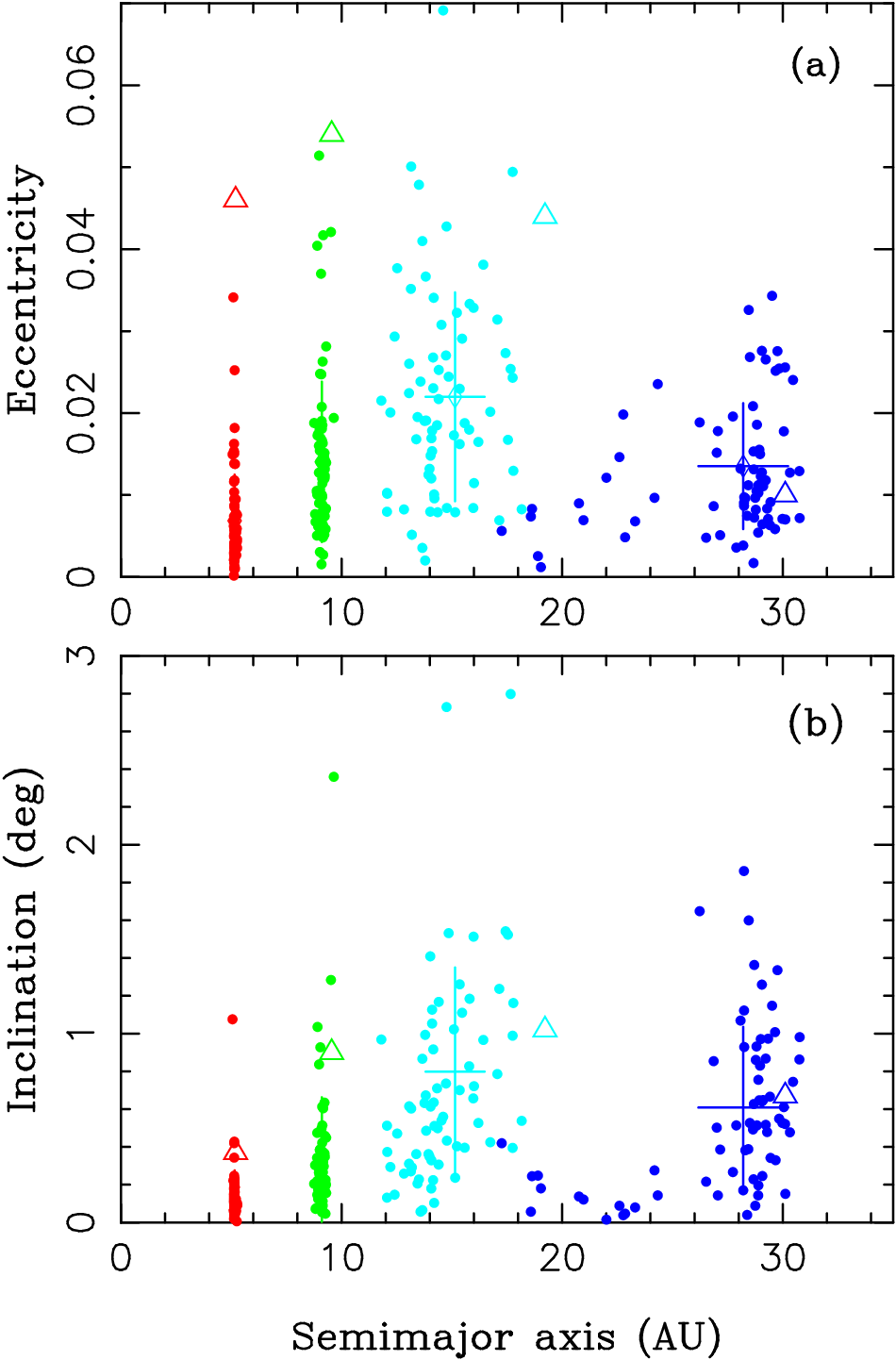}
\caption{Final orbits obtained in our simulations with four planets started in the (2:1,3:2,3:2) 
resonant chain, and $M_{\rm disk}=35$ M$_{\rm Earth}$. See the caption of Fig.
\ref{sum1} for the description of orbital parameters shown here.}
% job100/p4r21c2/m2d1_done
\label{sum2}
\end{figure}

The case with the 2:1 resonance between Jupiter and Saturn showed behavior that was reminiscent to that of the ONM 
(Tsiganis et al. 2005). The success rate for A and B was large, but Jupiter and Saturn migrated smoothly and 
criterion D was not satisfied. Also, because of the lack of an adequate excitation agent, Jupiter ended up with 
orbital eccentricity that was too low (Fig. \ref{sum2}), and C was not satisfied as well. In addition, Uranus 
typically reached $a_{\rm Ura}<18$ AU, well inside its present orbit. In the ONM (Tsiganis et al. 2005), the inner ice 
giant was started far beyond the 2:1 resonance with Saturn, which helped Uranus to reach 19.2 AU. 
\subsection{Case with 5 Planets}
Following Nesvorn\'y (2011) and Batygin et al. (2012) we considered the case with five giant planets as an 
interesting solution to some of the problems discussed in the previous section. The main advantage of the
five planet case is that the system can afford to lose a planet. This resolves, to a degree, the conflicting 
requirements for a significant excitation of $e_{55}$, which requires strong scattering encounters between
Jupiter and ice giants, and constraints A and B.   

We performed 4050 integrations with five planets for 19 different resonant chains. We tested different masses of 
the fifth planet and different initial orbits. We found that the best results were obtained when
the fifth planet had mass comparable to that of Uranus/Neptune, and was placed between the orbits of Saturn 
and Uranus. We discuss on this case below. The case in which the fifth planet was given a large mass typically 
resulted in a violent instability and ejection of Uranus and/or Neptune from the system. If the fifth planet
was given a lower mass and orbit beyond that of Neptune, the inner ice giant got ejected during the instability 
thus leaving a final system with incorrect masses. This later case bears similarities to the four-planet case 
discussed in Sect. 5.1.

\subsubsection{(3:2,3:2,4:3,5:4) and Alike}

Morbidelli et al. (2007) found one system with five planets that remained dynamically stable over 1 Gyr (in
absence of perturbations from the planetesimal disk).\footnote{In the published article, Morbidelli et al. 
(2007) claimed that the five-planet configuration was unstable. Unfortunately, to test the stability, Morbidelli 
et al. (2007) used the original version of {\it SyMBA} (Duncan et al. 1998), which had a problem with integrating 
very compact planetary configurations for long periods of time. After fixing this problem, Levison et al. (2011) 
found that the five-planet configuration identified by Morbidelli et al. (2007) was stable.} The five planets 
were in the (3:2,3:2,4:3,5:4) resonant chain. The extra ice giant had mass similar to that of Uranus/Neptune. The 
orbital eccentricities were 0.004, 0.02, 0.08, 0.035, and 0.01 (from the inner to outer planet).  Levison et al. 
(2011) showed that in such a case, where the inner ice giant had the largest eccentricity, it can be expected that 
the (late) instability was triggered by the inner ice giant (i.e., ${\cal B}(1)$ in notation introduced in Sect. 2). 
We discuss this case first and get to different trigger mechanisms and resonant chains later. 

We found that the low-mass disks with $M_{\rm disk} \lesssim 20$ M$_{\rm Earth}$ did not work, because the 
system frequently lost two or more planets, ending with three or less. Specifically, only $\sim$10\% of 
the simulations with $M_{\rm disk} = 20$ M$_{\rm Earth}$ ended up with four planets, but the orbits were all 
wrong. This is similar to what happened in the four-planet case described in Sect. 5.1.1 for 
$M_{\rm disk} \lesssim 35$ M$_{\rm Earth}$. The instability is simply too violent in this case to be contained 
by a low-mass planetesimal disk. Very massive planetesimal disks with $M_{\rm disk} > 50$ M$_{\rm Earth}$ did 
not work as well mainly because too much mass was processed through the Jupiter/Saturn region, leading to 
excessive migration of Jupiter and Saturn, and incorrect final $P_{\rm Sat}/P_{\rm Jup}>3$.

Intermediate disk masses, $M_{\rm disk} = 35$ M$_{\rm Earth}$ and 50 M$_{\rm Earth}$, were more promising.
With $\Delta =1$ AU (recall that $\Delta$ denotes the initial radial separation between the outer ice giant and inner edge of 
the planetesimal disk; see Sect. 2), these cases showed the 13\% and 37\% success rates for A, and 3\% and 23\% for B, 
respectively. These fractions are pretty much independent of the instability trigger. The larger success 
rate for $M_{\rm disk} = 50$ M$_{\rm Earth}$ than for $M_{\rm disk} = 35$ M$_{\rm Earth}$ reflects the stabilizing 
effect of the planetesimal disk, which increases with its mass. The success rate tends to drop with 
increasing $\Delta$, but this trend is not always clear. For example, the success rate for A drops from 
37\% to 23\% when $\Delta$ is increased to 3.5 AU for $M_{\rm disk} = 50$ M$_{\rm Earth}$ (similarly, B drops 
from 23\% to 10\%), while the percentages for $\Delta = 1$ AU and 3.5 AU are similar for $M_{\rm disk} = 35$ 
M$_{\rm Earth}$.

\begin{figure}
\epsscale{0.6}
 \plotone{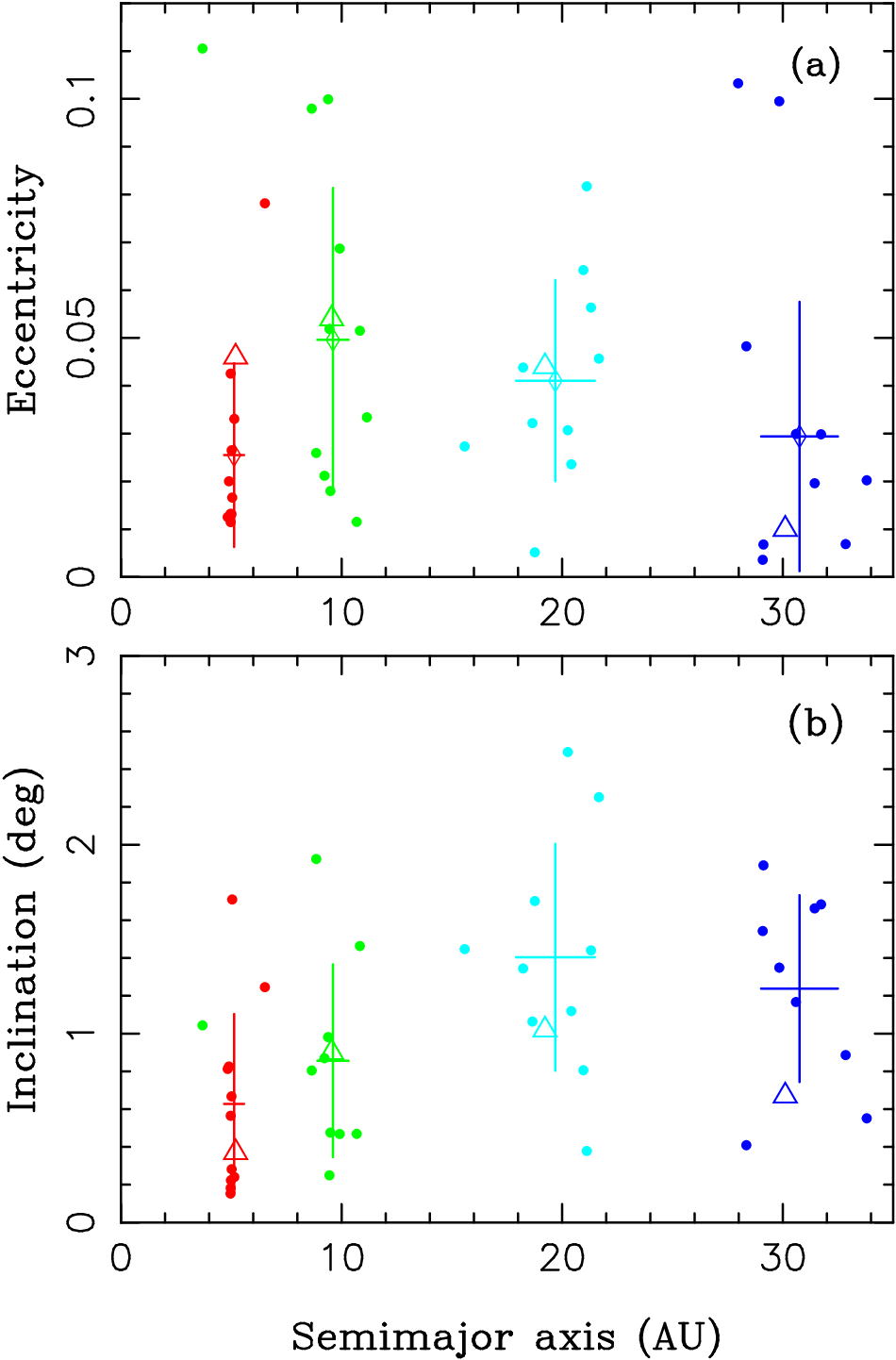}
\caption{Final orbits obtained in our simulations with five planets started in the (3:2,3:2,4:3,5:4) 
resonant chain, $M_{\rm disk}=50$ M$_{\rm Earth}$ and $\Delta=1$ AU. See the caption of Fig.
\ref{sum1} for the description of orbital parameters shown here. }
% p5/r32c1/m3d1/
\label{sum3}
\end{figure}

\begin{figure}
\epsscale{1.0}
\plotone{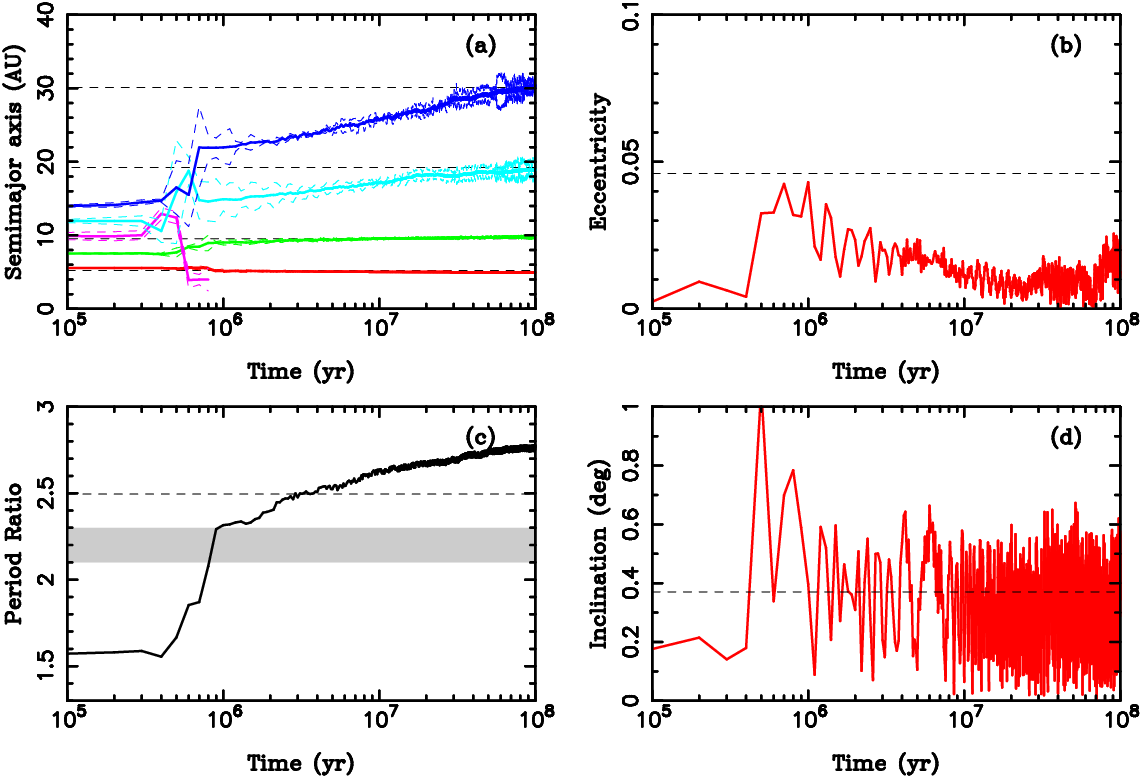}
\caption{Orbit histories of the giant planets in a simulation with five initial planets.
See the caption of Fig. \ref{case1} for the description of orbital parameters shown here. 
The five planets were started in the (3:2,3:2,4:3,5:4) resonant chain, $M_{\rm disk}=50$ 
M$_{\rm Earth}$ and $\Delta=1$ AU. The fifth planet was ejected at $t=0.8$ Myr after the start 
of the simulation.
% job number 13 from p5/r32c1/m3d2
}
\label{case4}
\end{figure}

Figure \ref{sum3} shows the final systems for $M_{\rm disk} = 50$ M$_{\rm Earth}$. This result is encouraging. 
The distributions show a larger range of values than what we got in the four-planet case, because the 
interaction of five planets is more complex and leads to a larger variety of results. Neptune's model orbit 
tends to be slightly more excited than the real one, but this difference is small and probably not fundamental 
in that it could be resolved by tuning the parameters. On the positive side, some of the simulations that 
satisfied A and B are now also showing signs of success for the criterion D (Fig. \ref{case4}). This has not 
happened in the four-planet case at all.

Still, the success rate for the criterion D is only marginal, mainly because it is intrinsically difficult to jump 
from $P_{\rm Sat}/P_{\rm Jup}=1.5$ to $>$2.3, and end up below 2.5. First, as the jump must be rather large, 
it is hard find cases where $P_{\rm Sat}/P_{\rm Jup}$ is larger than 2.3 and smaller than 2.5 after the jump. 
This is statistically unlikely (3.3\% averaging over the first seven lines of Table 6) 
given the spread of possible jump amplitudes for different encounter geometries. 
Second, there is a problem with the {\it residual} migration if the planetesimal disk is initially massive. In 
such cases, even if $2.3<P_{\rm Sat}/P_{\rm Jup}<2.5$ after the jump, there is still a significant mass in the 
planetesimal disk that needs to be processed through the Jupiter/Saturn zone. Consequently, $P_{\rm Sat}/P_{\rm Jup}$ 
keeps increasing after the jump and reaches $>2.5$ (for $M_{\rm disk} \gtrsim 50$~M$_{\rm Earth}$; e.g., Fig. \ref{case4}c). 
The residual migration is difficult to avoid unless the planetesimal disk had a relatively low mass to start with. 

An additional problem with the case discussed here is that $e_{55}$ that we obtained in the simulations was nearly 
always too small. This happened because even if $e_{\rm Jup}$ was kicked by the ejected planet, it was quickly damped 
at later times by secular friction from the planetesimal disk (see, e.g., Fig. \ref{case4}b). We tested several possible 
solutions to this problem. Disks with $M_{\rm disk} \lesssim 35$ M$_{\rm Earth}$ did not work for the (3:2,3:2,4:3,5:4) 
resonant chain because two or more planets were lost too often, end even if four planets survived in the end their
orbits were too wild (B satisfied in 5-10\% of cases only).  

We tested dynamically excited planetesimal disks. The disk excitation reduces the disk's ability to damp $e_{\rm Jup}$ 
and may thus lead to better results. The disk could have been excited by large bodies that formed in it, as evidenced
by Pluto-sized bodies in the present Kuiper belt. In addition, the disk could have been excited near its inner 
edge by orbital resonances with the outer ice giants. In both cases, the disk planetesimals were distributed 
according to the Rayleigh distribution. We used $\langle e \rangle = 0.15$ and $\langle i \rangle = 0.075$ (case E1),
and $\langle e \rangle = 0.1$ and  $\langle i \rangle = 0$ (case E2), where brackets denote the mean values. 
Case E1 mimicked the excitation expected from the large bodies in the disk. Case E2 would be more appropriate for 
the orbital resonances that do not generally excite inclinations. 

Recall that for the cold planetesimal disk we used a minimal separation of $\Delta\geq 0.5$ AU between the outer 
planet's orbit and semimajor axis of disk particles (see discussion in Sect. 2). Now that we deal with excited 
disks with $e$ up to 0.15, we used $\Delta\geq3.5$ AU for consistency, so that the minimal physical distance 
between planetesimals and planets was larger than 0.5 AU. 

With $M_{\rm disk} = 50$ M$_{\rm Earth}$, case E1 gave $A=30\%$ and $B=10$\%, 
which was comparable to what we obtained for this setup with a dynamically cold disk. Interestingly, there was one 
simulation (out of 30) that matched C as well ($e_{55}=0.032$). Case E2 gave $A=26\%$ and $B=7$\%, and also one 
simulation where the criterion C was satisfied ($e_{55}=0.026$). Conversely, the excited disks with $M_{\rm disk}=35$ 
M$_{\rm Earth}$ did not show any case where C would be satisfied. We conclude that the disk excitation may help, but 
the simulations matching C are still disturbingly rare.
 
We studied the dependence of the results on the mass of the fifth planet. We found that decreasing the fifth planet's mass helps 
to boost the success rate for A and B. For example, with (3:2,3:2,4:3,5:4), $M_{\rm disk} = 50$ M$_{\rm Earth}$, ${\cal B}(1)$, 
and the mass of the inner ice giant $M_{\rm Ice} = 0.5$ M$_{\rm Nep}$, the success rate in matching A and B was 
63\% and 30\%, while it was 37\% and 23\% with $M_{\rm Ice} = M_{\rm Nep}$. In all studied cases with 
$M_{\rm Ice} = 0.5$ M$_{\rm Nep}$, the small inner planet was removed (83\% ejected and 17\% collided with 
Jupiter).

On the downside, we found that the jump of $P_{\rm Sat}/P_{\rm Jup}$ and excitation of $e_{\rm Jup}$ with $M_{\rm Ice} 
< 1$ M$_{\rm Nep}$ were smaller than with $M_{\rm Ice} \simeq$ M$_{\rm Nep}$, which compromised the success rate for
C and D. In fact, $e_{\rm Jup}$ was excited in most of the runs with  $M_{\rm Ice} = 0.5$ M$_{\rm Nep}$ not by
the scattering encounter itself, but rather when Jupiter and Saturn crossed the 2:1 resonance. As we discussed 
Sect. 5.1, the resonant crossing has only a minor effect on $e_{55}$.

Increasing the mass of the inner ice giant did not help either, because the system became violently chaotic during
the instability and typically lost two or more planets. In addition, in many cases where only one planet was ejected, 
the lost planet was one of the outer ice giants. For example, with $M_{\rm disk} = 50$ M$_{\rm Earth}$, ${\cal B}(1)$
and $M_{\rm Ice} = 2 M_{\rm Nep}$ we found that $A=15$\% and $B=3$\% (when only the removal of the massive inner planet
was considered). We conclude that the problems with the low success rate of C and D cannot be resolved by changing 
the mass of the fifth planet.

The problems discussed above may be related to the fact that the ice giants' orbits were closely packed together
for (3:2,3:2,4:3,5:4), perhaps too closely. We found that the other compact resonant chains investigated here, such 
as (3:2,3:2,4:3,4:3) or even (3:2,3:2,3:2,4:3), behave in very much the same way as (3:2,3:2,4:3,5:4), giving way 
to violent instabilities that fail to satisfy the constraints. We therefore decided to focus on more relaxed resonant 
chains. We discuss the results obtained with (3:2,3:2,3:2,3:2) first. These results were more encouraging.

\subsubsection{Relaxed Chains with the 3:2 Jupiter-Saturn Resonance}

\begin{figure}
\epsscale{1.0}
\plotone{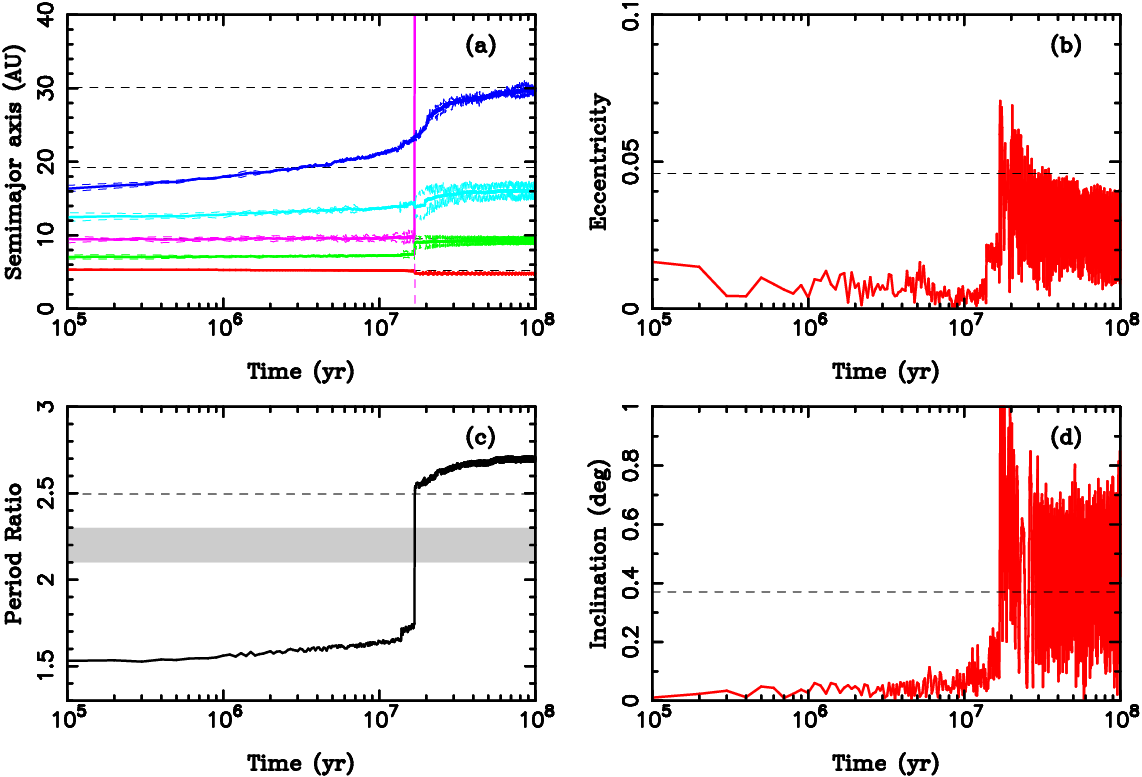}
\caption{Orbit histories of the giant planets in a simulation with five initial planets. 
See the caption of Fig. \ref{case1} for the description of orbital parameters shown here. 
The five planets were started in the (3:2,3:2,3:2,3:2) resonant chain, $M_{\rm disk}=35$ M$_{\rm Earth}$
and ${\cal B}(1)$. 
The fifth planet was ejected at $t=17$ Myr after the start of the simulation.}
% job number 13 from job100,p5r32c1/m2d1
\label{case5}
\end{figure}

The (3:2,3:2,3:2,3:2) resonant chain leads to a different mode of instability, if $\Delta < 2$~AU. Figure \ref{case5} 
illustrates a case with $M_{\rm disk} = 35$ M$_{\rm Earth}$, ${\cal B}(1)$ and $\Delta = 1.5$ AU. Unlike in all cases 
discussed so far, breaking the resonant lock of the inner ice giant did not result in an immediate instability.
Instead, Neptune migrated deep into the planetesimal disk {\it before} any scattering encounters between planets occurred. 
As Neptune and Uranus scattered planetesimals into the Jupiter/Saturn zone, Jupiter and Saturn underwent divergent 
migration and left the 3:2 resonance. Their eccentricities were then excited by the the 5:3 resonance crossing, and 
the instability happened shortly after (16.8 Myr after the start of the simulation). 

The epoch of planetary encounters was brief, lasted from 16.816 Myr to 16.9 Myr at which point the fifth planet was 
ejected from the Solar System (escape speed $V_\infty=0.74$~km~s$^{-1}$).\footnote{Typically, $V_\infty=0.5$-2 km s$^{-1}$.} 
All planets participated in the encounters, as required for capture of the irregular satellites,\footnote{We do not 
define planetary encounters as a separate criterion, because constraints C and D require planetary encounters and are 
generally much more restrictive.} but all these encounters involved the fifth planet only. None of the surviving giant 
planets had an encounter with another surviving giant. In total, 282 encounters happened in this case, where an encounter 
is defined here by the condition that the Hill spheres of the two planets at least partially overlap.

Now, the case illustrated in Fig. \ref{case5} simultaneously matched {\it all} our criteria. Amplitude $e_{55}$ was 
excited by the ejection of the fifth planet by Jupiter and was not damped too much during the following evolution, 
because the planetesimal disk had a relatively low initial mass, and was partially disrupted by Neptune prior to the 
excitation event. The final amplitudes were $e_{55}=0.024$ and $e_{56}=0.012$. Here, the amplitude $e_{55}$ was lower 
than real one, but in another simulation (with the same setup) that matched all our criteria as well, we obtained 
$e_{55}=0.041$ and $e_{56}=0.013$. 

\begin{figure}
\epsscale{1.0}
\plotone{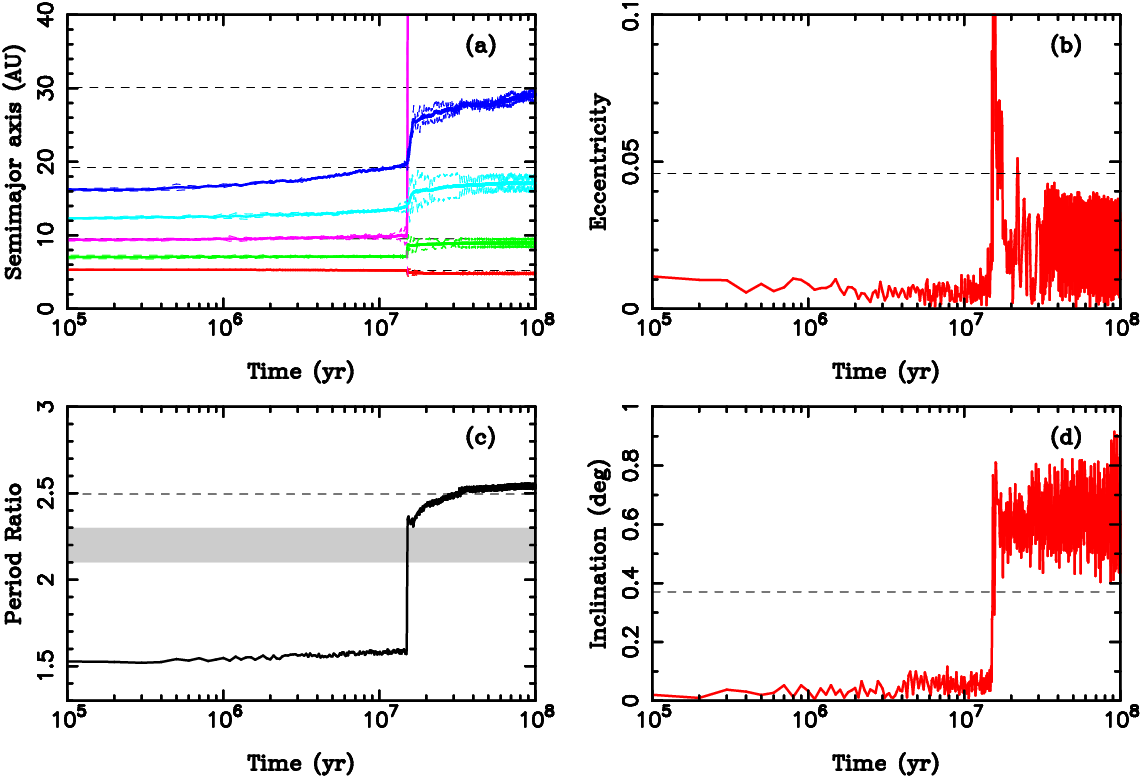}
\caption{Orbit histories of the giant planets in a simulation with five initial planets. 
See the caption of Fig. \ref{case1} for the description of orbital parameters shown here. 
The five planets were started in the (3:2,3:2,3:2,3:2) resonant chain, $M_{\rm disk}=35$ M$_{\rm Earth}$
and ${\cal B}(0)$ 
The fifth planet was ejected at $t=15.2$ Myr after the start of the simulation.}
% job number 3 from p5/r32c2/m2d0}
\label{case6}
\end{figure}

The scattering encounters with the fifth planet produced a jump of $P_{\rm Sat}/P_{\rm Jup}$ from 1.7 to 2.5, just as 
required to avoid the secular resonances with the terrestrial planets (Fig. \ref{case5}c). There was a small problem 
with the residual migration in this run, because $P_{\rm Sat}/P_{\rm Jup}$ continued to evolve past 2.5, while 
$P_{\rm Sat}/P_{\rm Jup}=2.49$ in the present Solar System, but we obtained better results in another simulation with 
the same setup (Figure \ref{case6}; although in this case $e_{55}$ was damped to 0.013). 

\begin{figure}
\epsscale{0.6}
\plotone{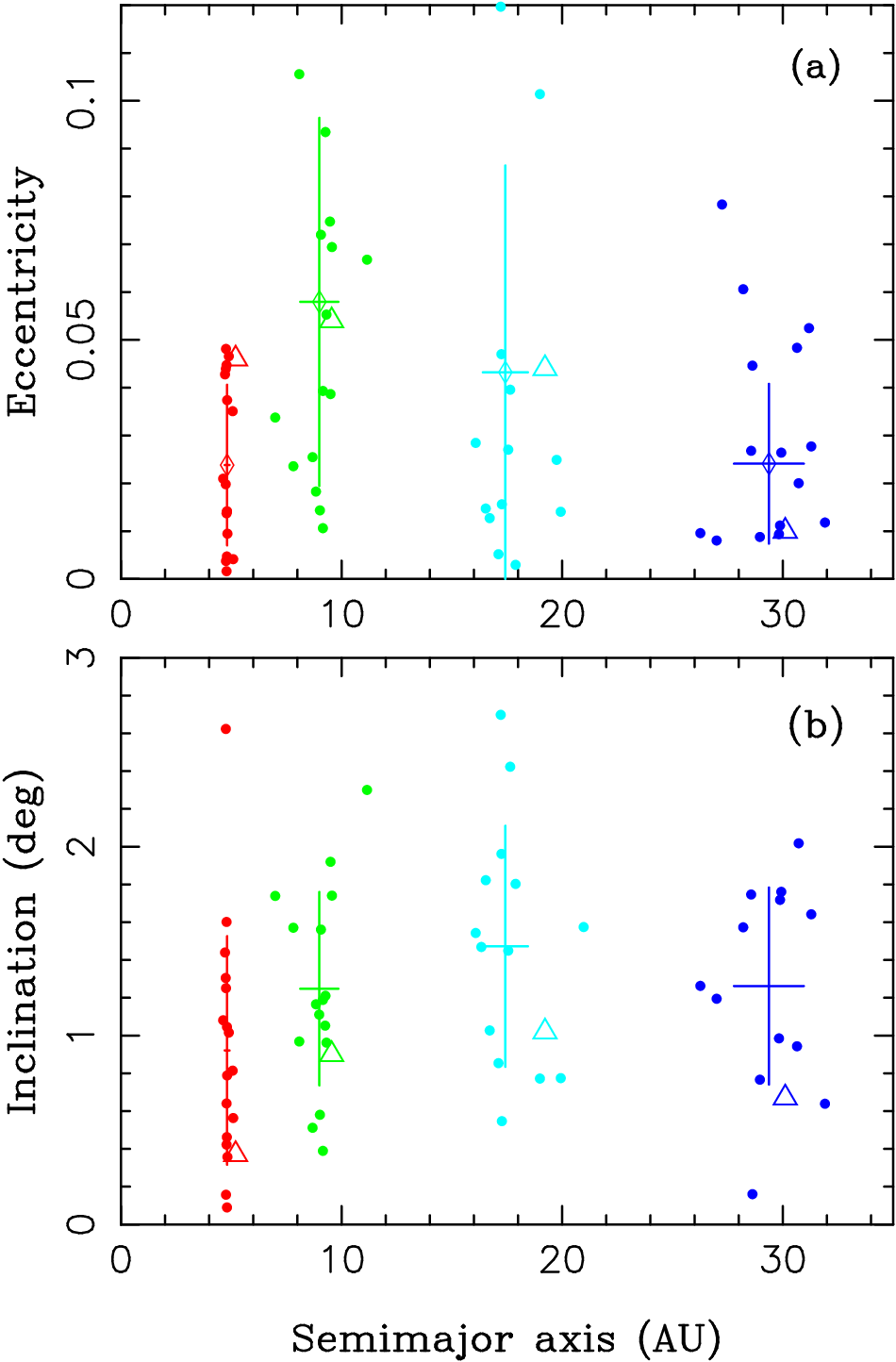}
\caption{Final orbits obtained in our simulations with five planets started in the (3:2,3:2,3:2,3:2) resonant 
chain, $M_{\rm disk}=35$ M$_{\rm Earth}$ and ${\cal B}(1)$.
See the caption of Fig. \ref{sum1} for the description of orbital parameters shown here.}
\label{sum4}
%job100/p5r32c1/m2d1
\end{figure}

The overall distribution of planetary orbits obtained in simulations with the (3:2,3:2,3:2,3:2) resonant chain, $M_{\rm disk} 
= 35$ M$_{\rm Earth}$, ${\cal B}(1)$ and $\Delta = 1.5$ AU is shown in Fig. \ref{sum4}. The overall success rate in matching 
the criteria A, B, C and D was 33\%, 16\%, 4\% and 8\%, respectively (note that matching C or D required that B was 
matched). The overall success rate for simultaneous matching of C {\it and} D was 4\%. This is the best result discussed 
so far. For a comparison, a similar setup with a more massive disk, $M_{\rm disk} = 50$ M$_{\rm Earth}$, gave to $A=30$\%,
$B=17$\%, $C=0$\% and $D=3$\%. The success for C dropped because $e_{55}$ was damped more efficiently with the massive 
planetesimal disk. A light disk with $M_{\rm disk} = 20$ M$_{\rm Earth}$ gave $A=20$\%, $B=7$\%, $C=3$\% and $D=0$\%.

The mode of instability discussed for (3:2,3:2,3:2,3:2) above may be problematic because the migration of Uranus and Neptune
prior to the instability may require that the inner edge of the planetesimal disk was initially close to the outer ice giant.
The edge may then need to be fine-tuned to generate the delay between the formation of the Solar System and the LHB. This is 
a problem similar to that of the original Nice model, which prompted Levison et al. (2011) to consider distant disks and a 
different trigger mechanism. Here we do not study the problem of delay in detail. Instead, given that the instability mode 
with prior migration of Neptune works much better than any other case, we pursued investigations into this instability mode
further.

We first tested several cases where the inner ice giant was placed the 2:1 resonance with Saturn ((3:2,2:1,2:1,3:2), 
(3:2,2:1,3:2,2:1), etc.). We found that these resonant chains did not work, because the orbits spread without
suffering any major instability, which left five planets behind. Unlocking the resonance of one of the ice giants did 
not destabilize the system as well, mainly because the inner ice giant was initially far from Saturn, and safe, even
if its orbit was not locked deep in the 2:1 resonance.  

Motivated by these results, we considered several resonant chains with the inner ice giant in the exterior 3:2 (or 4:3)
resonance with Saturn, {\it and} in the 2:1 resonance with the inner surviving ice giant (e.g., (3:2,3:2,2:1,2:1),
(3:2,4:3,2:1,2:1) and (3:2,3:2,2:1,3:2). This initial setup increases the likelihood that the inner ice giant gets 
ejected, because it starts relatively close to Saturn, and that the outer two ice giants survive, because there
is initially a large radial gap between them and the inner ice giant.

We found that the extended resonant chains such as (3:2,3:2,2:1,2:1) and (3:2,4:3,2:1,2:1) did not work because 
Neptune migrated beyond 30 AU, independently of the assumed mass and extension of the planetesimal disk. The 
explanation for this is clear. As in all simulations, Neptune scattered planetesimals outward and inward relative 
to its orbit. Many of those that were scattered outward were scattered to $a>30$ AU, because initial $a_{\rm Nep} 
\simeq 25$ AU for the extended resonant chains. The ones that were scattered inward must have been scattered 
relatively strongly to be handed to Uranus, because the orbits of Uranus and Neptune were radially spaced (due to 
the initial 2:1 resonance). 
As a result, Neptune migrated outward, being propelled by the strong scattering encounters, all the way into the 
cloud of planetesimals that it scattered beyond 30 AU. This occurred even if the planetesimal disk had low mass 
($M_{\rm disk} = 20$ M$_{\rm Earth}$) and/or was initially narrow (spanning, e.g., 26-27 AU). Also, even lighter 
disks with $M_{\rm disk} \lesssim 15$ M$_{\rm Earth}$ did not show much success for A.

\begin{figure}
\epsscale{0.6}
\plotone{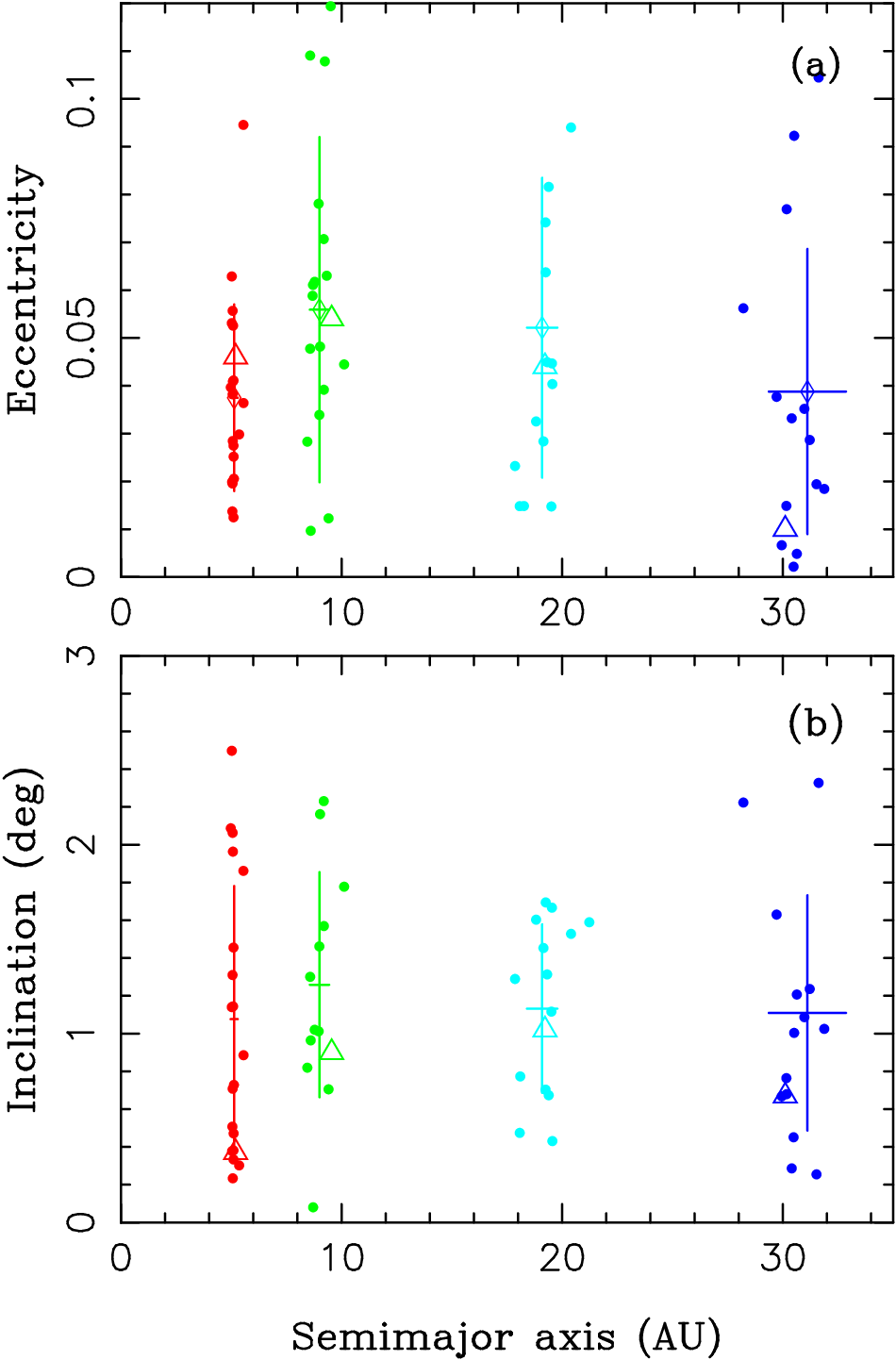}
\caption{Final orbits obtained in our simulations with five planets started in the (3:2,3:2,2:1,3:2) resonant 
chain, and $M_{\rm disk}=20$ M$_{\rm Earth}$. Results obtained with ${\cal B}(0)$ and ${\cal B}(1)$ for 
$\Delta=1$ AU were combined here. See the caption of Fig. \ref{sum1} for the description 
of orbital parameters shown here.}
\label{sum5}
%p5/r32c5b/m1d0 and m1d1
\end{figure}

Figure \ref{sum5} shows the distribution of final orbits for the (3:2,3:2,2:1,3:2) resonant chain and 
$M_{\rm disk} = 20$ M$_{\rm Earth}$. These results are more promising. As the distributions obtained with
${\cal B}(0)$ and ${\cal B}(1)$ did not differ (assuming $\Delta\simeq1$ AU), we combined them together in 
Fig.~\ref{sum5}. The combined success rate was $A=32$\%, $B=12$\%, $C=7$\%, $D=5$\% and $C\&D=5$\%. This was similar to the 
statistics obtained for (3:2,3:2,3:2,3:2) and $M_{\rm disk} = 35$ M$_{\rm Earth}$ (Fig.~\ref{sum4}), but here things were 
slightly better in detail. 

\begin{figure}
\epsscale{1.0}
\plotone{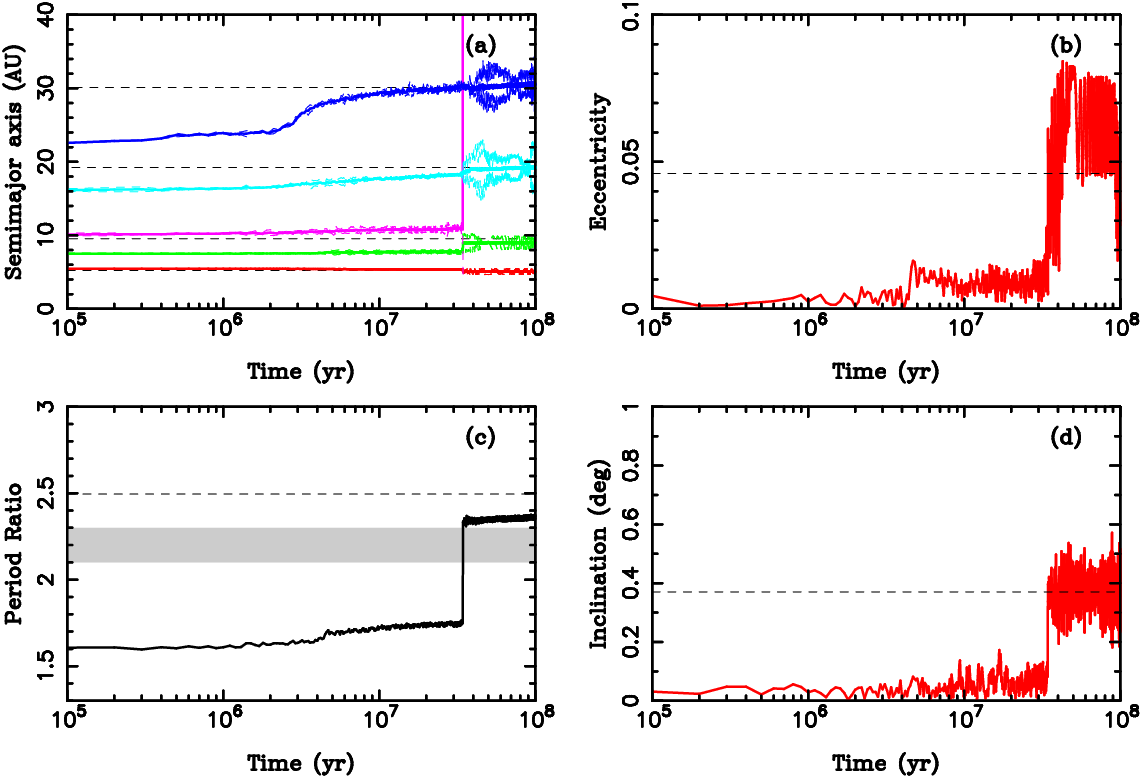}
\caption{Orbit histories of the giant planets in a simulation with five initial planets. 
See the caption of Fig. \ref{case1} for the description of orbital parameters shown here. 
The five planets were started in the (3:2,3:2,2:1,3:2) resonant chain, 
$M_{\rm disk}=20$ M$_{\rm Earth}$ and ${\cal B}(1)$. 
The fifth planet was ejected at $t=34.4$ Myr after the start of simulation.}
% job number 13 from p5/r32c5b/m1d1}
\label{case7}
\end{figure}

\begin{figure}
\epsscale{1.0}
\plotone{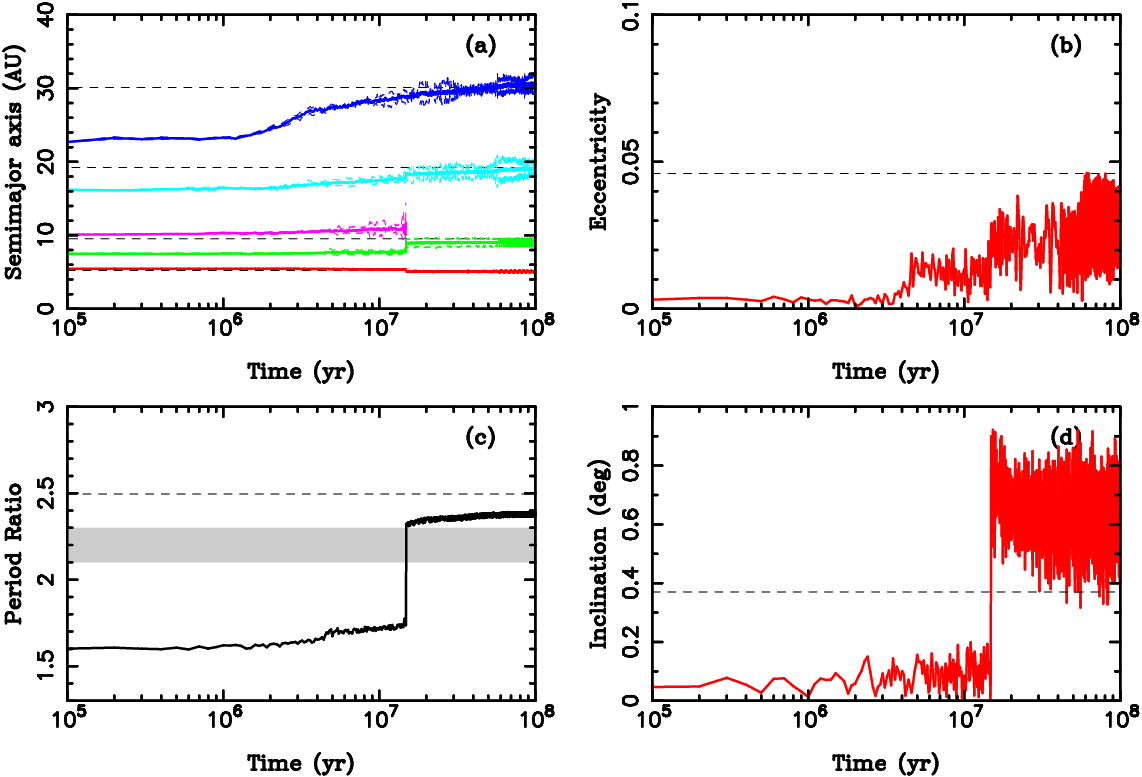}
\caption{Orbit histories of the giant planets in a simulation with five initial planets. 
See the caption of Fig. \ref{case1} for the description of orbital parameters shown here. 
The five planets were started in the (3:2,3:2,2:1,3:2) resonant chain, 
$M_{\rm disk}=20$ M$_{\rm Earth}$ and ${\cal B}(0)$.
The fifth planet was ejected at $t=14.9$ Myr after the start of the simulation.}
% job number 28 from p5/r32c5b/m1d0}
\label{case8}
\end{figure}

\begin{figure}
\epsscale{1.0}
\plotone{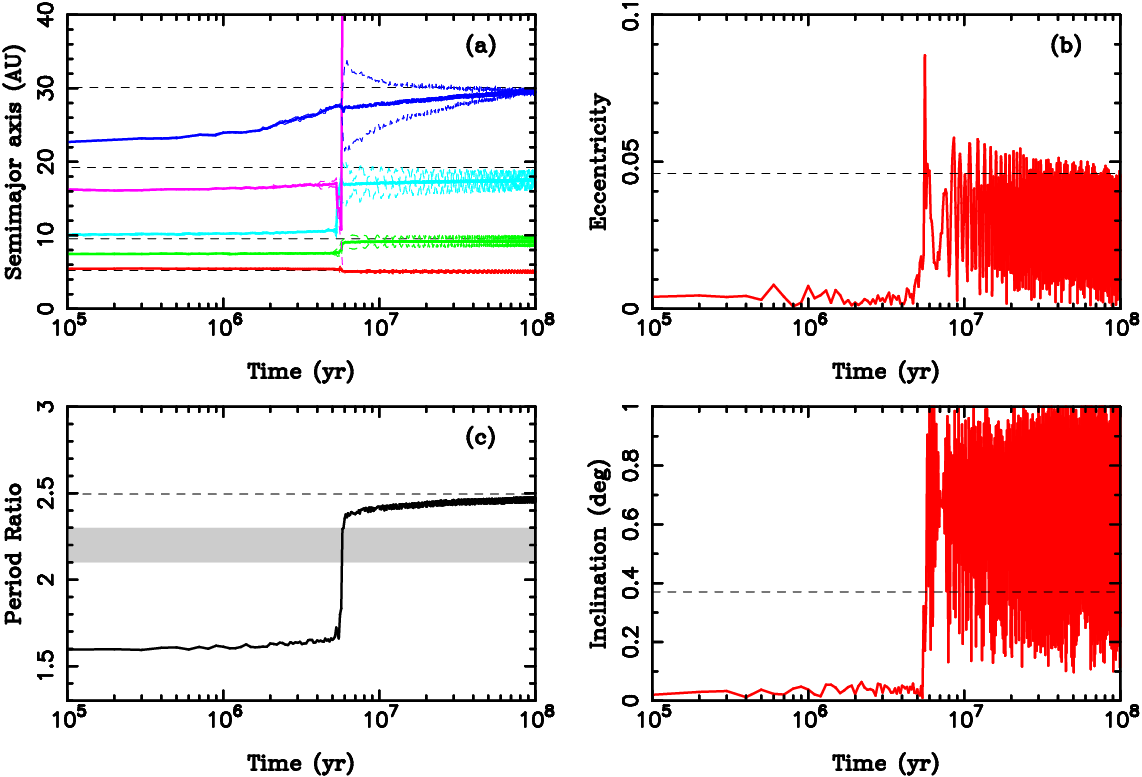}
\caption{Orbit histories of the giant planets in a simulation with five initial planets. 
See the caption of Fig. \ref{case1} for the description of orbital parameters shown here. 
The five planets were started in the (3:2,3:2,2:1,3:2) resonant chain, 
$M_{\rm disk}=20$ M$_{\rm Earth}$ and ${\cal B}(1)$.
The fifth planet was ejected at $t=6.1$ Myr after the start of the simulation.}
% job number 12 from p5/r32c5b_new/m1d1}
\label{case9}
\end{figure}

For example, the distribution of model $a_{\rm Ura}$ nicely matched the present semimajor axis of Uranus, while
the model $a_{\rm Ura}$ was slightly lower than required in Fig. \ref{sum4}. The model values of $e_{\rm Jup}$ 
obtained with (3:2,3:2,2:1,3:2) were almost perfect in that they closely surrounded the target 
$\bar{e}_{\rm Jup}=0.046$. Figures \ref{case7}, \ref{case8} and \ref{case9} show examples of the successful 
simulations. We illustrate these cases in detail because they were some of the best results obtained
in this work. 

The simulation in Fig. \ref{case7} ((3:2,3:2,2:1,3:2) and $M_{\rm disk} = 20$ M$_{\rm Earth}$) ended with $g_5=4.95$ 
arcsec yr$^{-1}$, $g_6=29.7$ arcsec yr$^{-1}$, $g_7=4.53$ arcsec yr$^{-1}$, and $g_8=0.52$ arcsec yr$^{-1}$. These values 
were a close match to those listed for the real planets in Table 2. The amplitudes of eccentricity modes were 
$e_{55}=0.049$, $e_{56}=0.033$, 
$e_{57}=0.024$, $e_{65}=0.041$, $e_{66}=0.042$, $e_{67}=0.003$, and $e_{77}=0.039$. The frequencies and amplitudes of
inclination modes were equally good. Neptune ended with an eccentricity that was about twice as large 
as its present value, but this was at least partly related to a minor mismatch in the simulation, because
Uranus and Neptune crossed the 2:1 resonance, which did not happened in reality. 

Figure \ref{case8} shows another case obtained with the same setup. 
In this case, $g_5=4.83$ arcsec yr$^{-1}$, $g_6=28.9$ arcsec yr$^{-1}$, $g_7=3.55$ arcsec yr$^{-1}$, 
and $g_8=0.64$ arcsec yr$^{-1}$. The amplitudes were $e_{55}=0.027$, $e_{56}=0.014$, $e_{57}=0.002$, $e_{65}=0.041$, 
$e_{66}=0.024$, $e_{67}=0.002$, $e_{77}=0.033$. Finally, the case illustrated in Fig. \ref{case9} had frequencies 
$g_5=4.66$ arcsec yr$^{-1}$, $g_6=28.1$ arcsec yr$^{-1}$, $g_7=3.45$ arcsec yr$^{-1}$, $g_8=0.61$ arcsec yr$^{-1}$,
and amplitudes $e_{55}=0.024$, $e_{56}=0.019$, $e_{57}=0.004$, $e_{65}=0.019$, $e_{66}=0.057$, $e_{67}=0.004$,
$e_{77}=0.040$, $e_{88}= 0.007$. This case differs with respect to the previous two in that the innermost ice
giant survived and evolved onto Uranus-like orbit, while the middle ice giant was ejected. The Neptune's model 
eccentricity was very good in this simulation. 

Using (3:2,3:2,2:1,3:2) and $M_{\rm disk} \simeq 20$ M$_{\rm Earth}$ we tested how the results
were affected by small changes (up to 50\%) in the mass of the fifth planet. We found that it was probably
not lower than 0.7 M$_{\rm Ura}$ because in the simulations with these low masses Jupiter (and Saturn) were
not kicked enough by the planet's ejection. Masses slightly larger than M$_{\rm Nep}$ showed lower success rates 
for A and B, but still worked relatively well for C and D. To summarize, we find that the fifth planet probably 
had mass in the Uranus/Neptune range, which is encouraging in that we do not need to invoke any special mass 
regime. 
\subsubsection{2:1 Jupiter-Saturn Resonance}
The initial conditions discussed in this section are probably academic, because the 2:1 resonance between 
Jupiter and Saturn is not favored by the standard model of the migration of planets in the protoplanetary gas disks, 
and their capture in resonances (see Sect. 3). We therefore discuss this case briefly, 
concentrating on the main differences with respect to the results obtained with more conservative assumptions. 

The five-planet case with the 2:1 resonance would require that the (ejected) inner ice giant had lower mass,
because when Jupiter and Saturn start in the 2:1 resonance, their period ratio needs to change by $\sim$0.5 
only (from 2 to 2.49), which requires a smaller perturbation. The best results were obtained with 
$M_{\rm Ice}=0.5$ M$_{\rm Nep}$ and $15 \lesssim M_{\rm disk} \lesssim 35$ M$_{\rm Earth}$. For example, with 
$M_{\rm disk}=20$ M$_{\rm Earth}$, we obtained a $\sim$50-70\% success for A and 20-40\% success for B. In addition,
because the required jump of $P_{\rm Sat}/P_{\rm Jup}$ is smaller and easier to achieve, the simulations also
show a large success for D (reaching 20\% in the most favorable cases).  

\begin{figure}
\epsscale{0.6}
\plotone{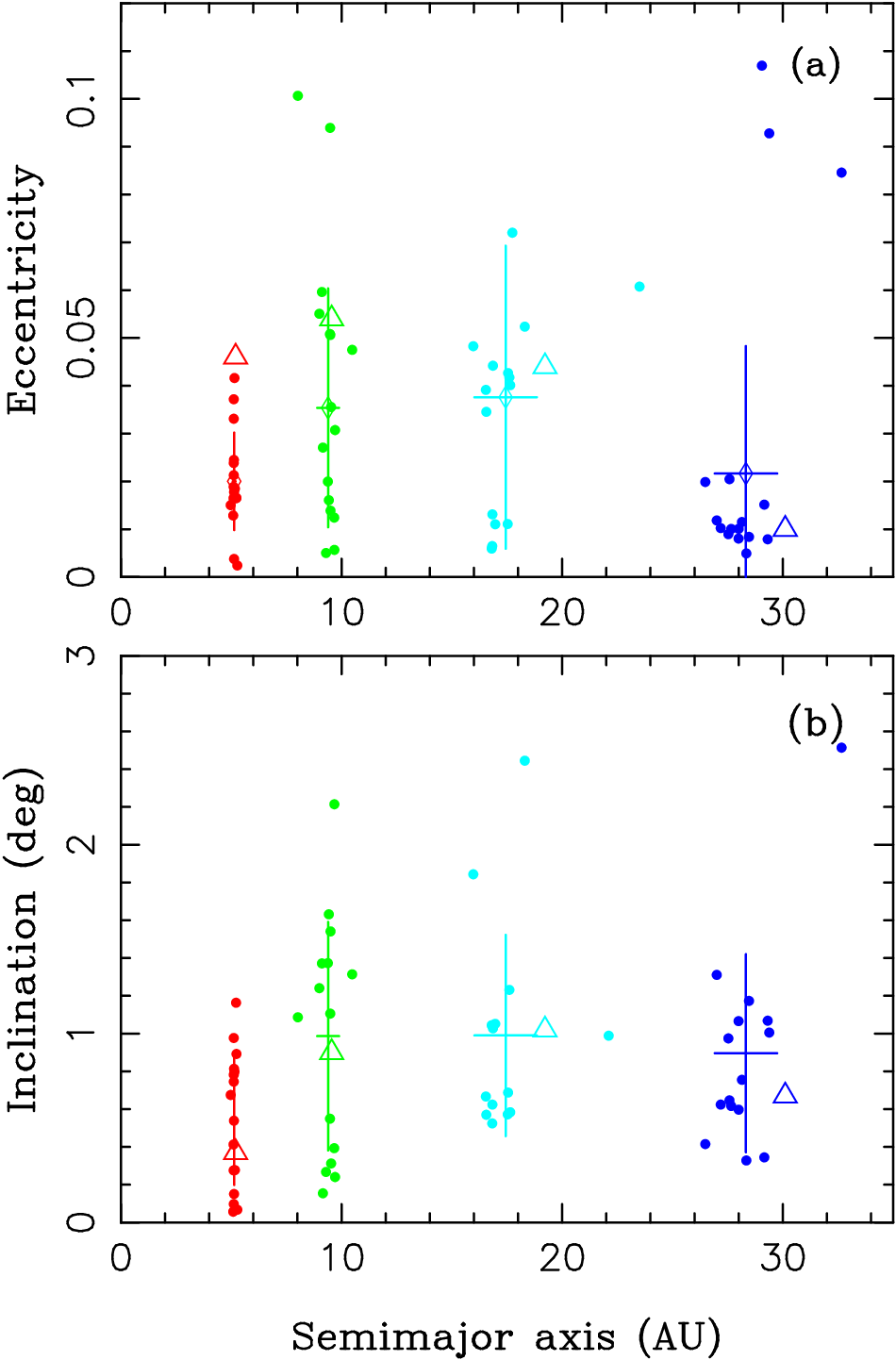}
\caption{Final orbits obtained in our simulations with five planets started in the (2:1,3:2,3:2,3:2) resonant 
chain, and $M_{\rm disk}=20$ M$_{\rm Earth}$. See the caption of Fig. \ref{sum1} for the description 
of orbital parameters shown here.}
\label{sum6}
%p5/r21c1_half/m1d1
\end{figure}

Figure \ref{sum6} illustrates this case. The model semimajor axis of Uranus and Neptune were a bit smaller than 
what would be ideal, but this should easily be corrected by extending the disk slightly beyond 30 AU (we used
$r_{\rm out}=30$ AU). All else looked great, {\it except} that the Jupiter's eccentricity was generally too 
small and violated constraint C. We found only two simulations out of more than 1,000, where the criterion C 
was satisfied. 

\begin{figure}
\epsscale{1.0}
\plotone{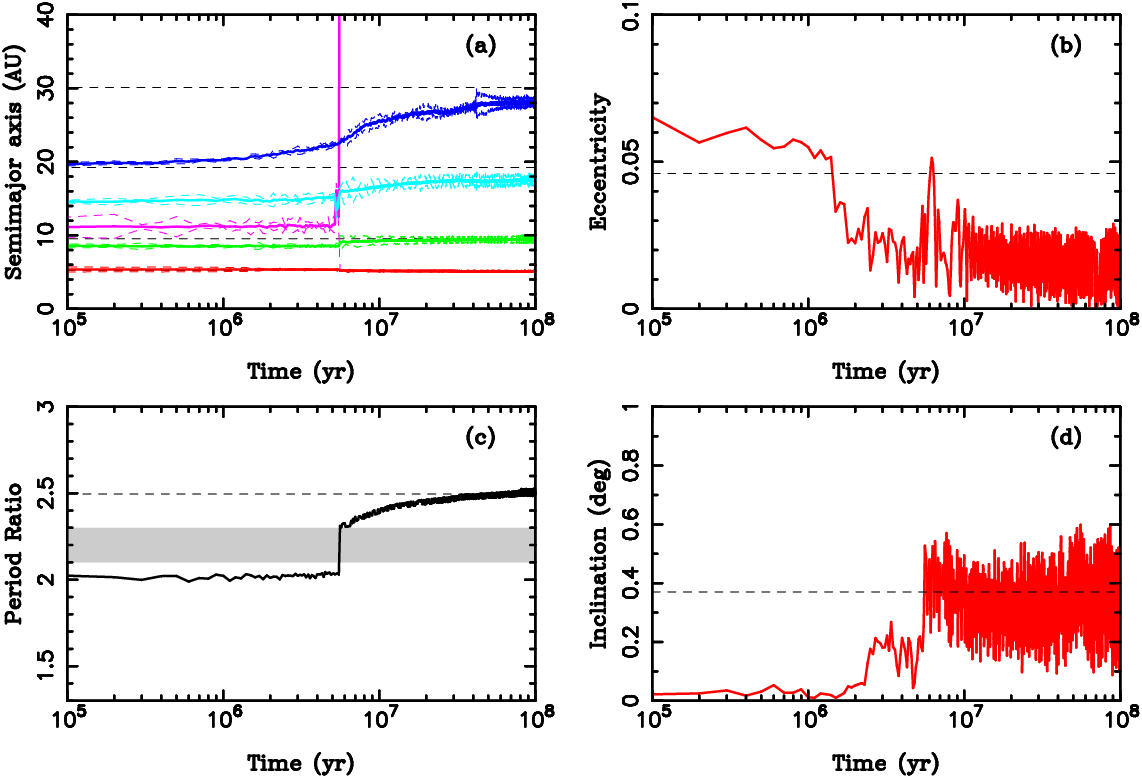}
\caption{Orbit histories of the giant planets in a simulation with five initial planets. 
See the caption of Fig. \ref{case1} for the description of orbital parameters shown here. 
The five planets were started in the (2:1,3:2,3:2,3:2) resonant chain, and $M_{\rm disk}=20$ M$_{\rm Earth}$. 
The fifth planet was ejected at $t=5.6$ Myr after the start of the simulation.}
% job number 17 from p5/r21c1_half/m1d1
\label{case10}
\end{figure}

This problem arises because while a relatively low-mass ice giant is required to produce the needed small jump 
of $P_{\rm Sat}/P_{\rm Jup}$, the same low-mass planet is generally incapable of exciting $e_{55}$ enough during 
encounters. In addition, the problem with secular friction discussed in the previous sections is also
apparent here. To show this clearly, we adjusted the initial system so that Jupiter had a substantial eccentricity 
initially (Fig. \ref{case10}). This did not helped at all because the initial eccentricity was quickly damped. 

The problem with matching C in the case with five planets and 2:1 resonance between Jupiter and Saturn is difficult 
to avoid (at least we were not able to resolve this problem with 1,000+ simulations). Thus, even if the success rate 
for A, B and D was promisingly large, we are not overly optimistic about this case. A more thorough testing will
be required to reach a more definitive conclusion.\footnote{The discussion presented here was based on the overall 
synthesis of our simulation results. Note that it can be misleading to compare two specific simulation sets in 
Table 6. For example, we have just been lucky to find one case (out of 30) for (2:1,3:2,3:2,3:2), $M_{\rm disk}=20$ 
M$_{\rm Earth}$ and ${\cal B}(1)$ for which the criterion C was satisfied. In contrast, the results for the 
3:2 Jupiter-Saturn resonance were consistently good, showing $>$1\% success rate for matching all criteria 
simultaneously.}

\subsection{Case with 6 Planets}
We performed 1290 simulations in total for six initial planets and nine different resonant chains. Given that
the number of parameters in the six-planet case is much larger than the one with four or five planets, we are
not confident that we sampled parameter space exhaustively enough to make detailed conclusions. Still, the 
six-planet case is very interesting because we were able to obtain good results even after a relatively 
small number of trials. 

\begin{figure}
\epsscale{0.6}
\plotone{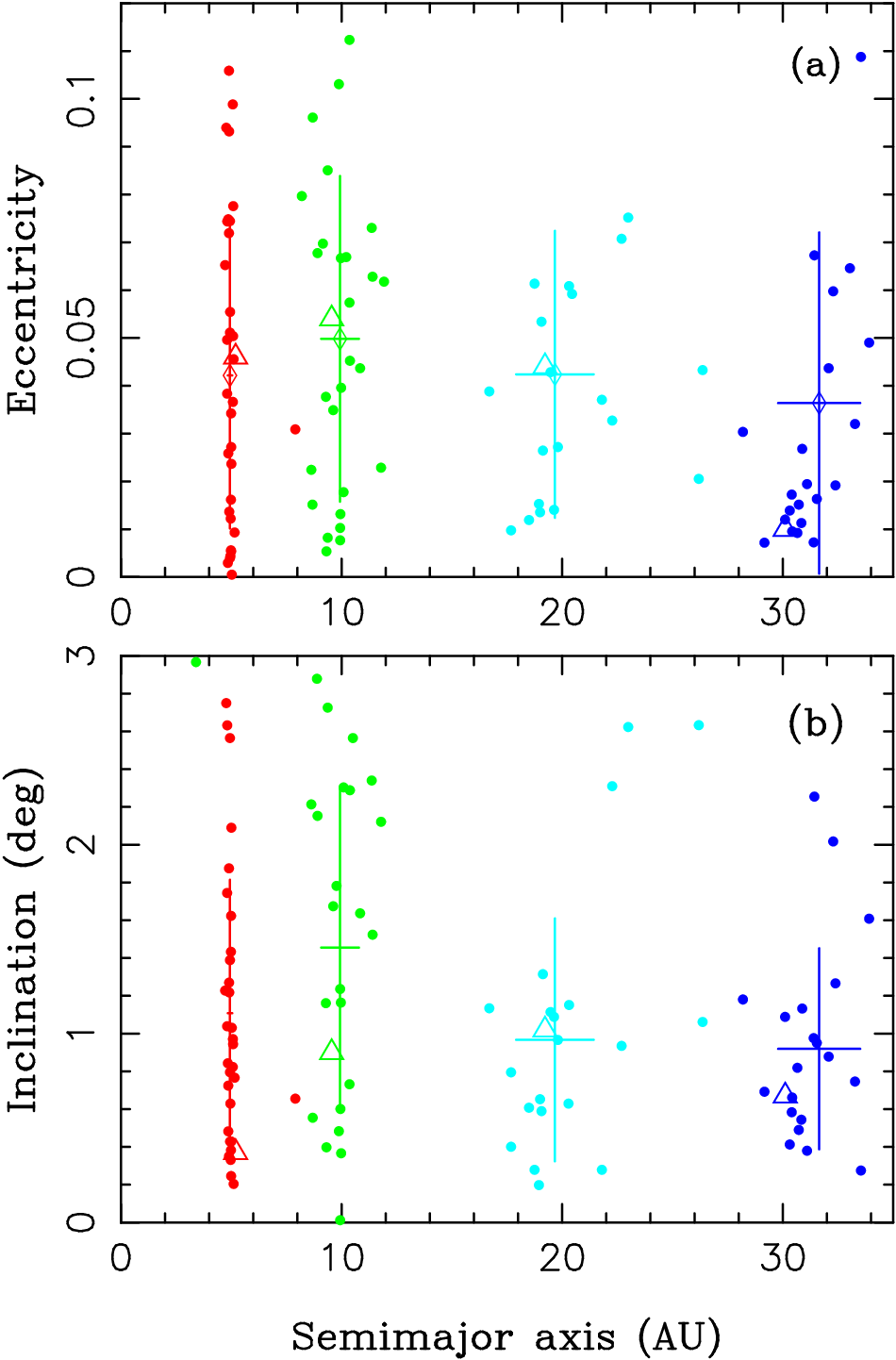}
\caption{Final orbits obtained in our simulations with six planets started in the (3:2,4:3,3:2,3:2,3:2) resonant 
chain, and $M_{\rm disk}=20$ M$_{\rm Earth}$. See the caption of Fig. \ref{sum1} for the description 
of orbital parameters shown here.}
\label{sum7}
%job100/p6r32c2/m1d3
\end{figure}

Figure \ref{sum7} illustrates one of the most promising six-planet results. This result was obtained for the 
(3:2,4:3,3:2,3:2,3:2) resonant chain and $M_{\rm disk}=20$ M$_{\rm Earth}$. The two extra ice giants were placed
between Saturn and the inner surviving ice giant, and were given masses $M_{\rm Ice1}=M_{\rm Ice2}=0.5$~M$_{\rm Nep}$
(the six-planet case with $M_{\rm Ice1}=M_{\rm Ice2}=1$~M$_{\rm Nep}$ does not work, because the instability is too 
violent).

The results show a large spread, because the six planets have complex interactions during the instability. The 
distributions of model orbits nicely overlap with the real orbits. Even the details match. For example, 
most simulations ended with $e_{\rm N}\simeq0.01$, just as needed. Given our previous experience with the 
simulations of instability in the four- and five-planet cases, the agreement in Fig. \ref{sum7} is remarkable. 
Because of the larger spread, however, the success rate was lower than that for our best five-planet simulations. In 
the specific case discussed above, we obtained $A=30$\%, $B=10$\%, $C=3$\%, $D=3$\% and $C\&D=2$\%.

\begin{figure}
\epsscale{1.0}
\plotone{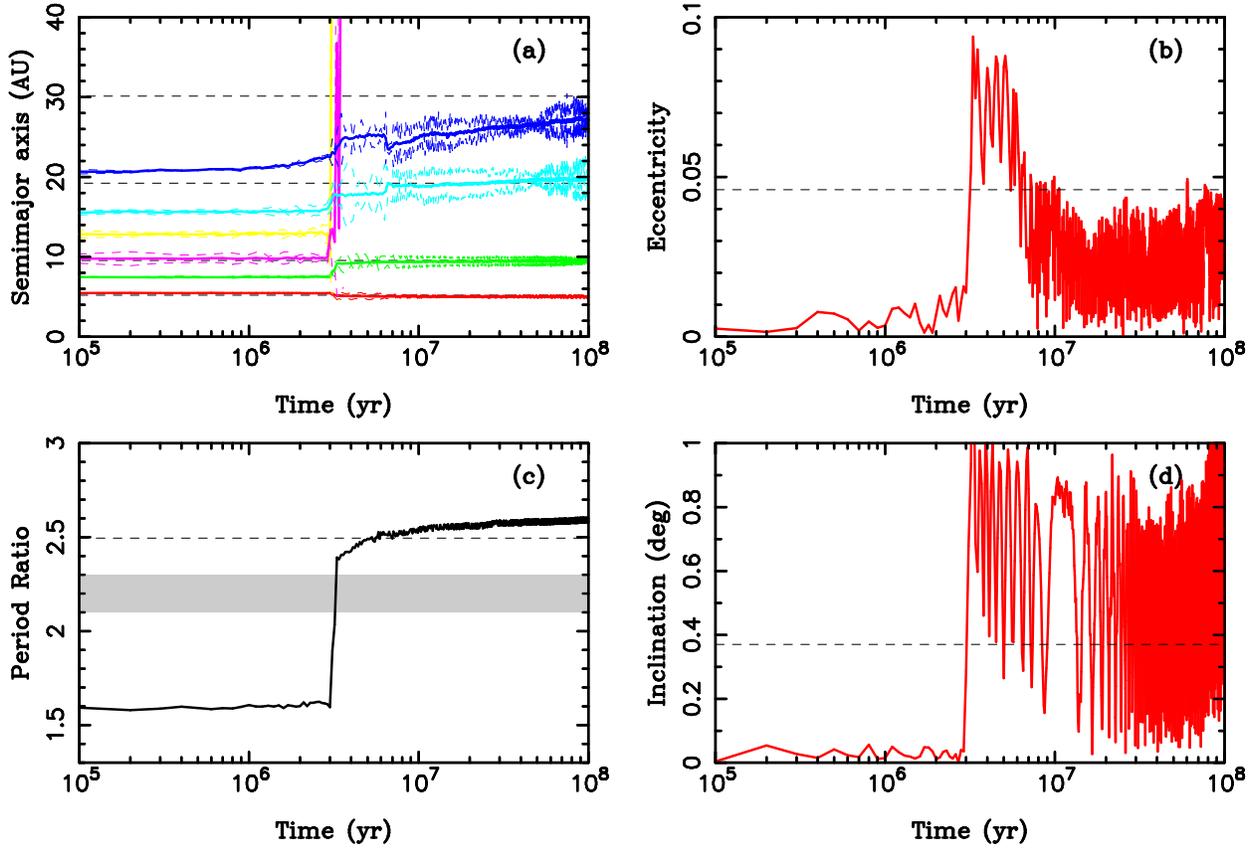}
\caption{Orbit histories of the giant planets in a simulation with six initial planets. 
See the caption of Fig. \ref{case1} for the description of orbital parameters shown here. 
The six planets were started in the (3:2,4:3,3:2,3:2,3:2) resonant chain, and $M_{\rm disk}=20$ M$_{\rm Earth}$. 
The fifth and sixth planet were ejected at $t=3.18$ and 3.65 Myr after the start of simulation (yellow 
and purple lines in (a)).}
% job number 11 from p6/r32c1/m1d3
\label{case11}
\end{figure}

The instability typically occurred in two steps, corresponding to the ejection of the two planets. Sometimes, as 
in Fig. \ref{case11}, the ejection of the two planets was nearly simultaneous, but most of the times there was
a significant delay between ejections.  This was useful because the first planet's ejection partially disrupted 
the planetesimal disk and reduced its capability to damp $e_{\rm 55}$, which was then excited by the second planet's 
ejection. While this mode of instability can be important, we would need to increase the statistics ($>$100 
simulations for each initial condition) to be able to properly resolve the small success fractions in the 
six-planet case. 
\section{Discussion}
One of the main results discussed in this paper is that the Solar System instability with five initial planets
tends to give better results then the instability with four initial planets. Batygin et al. (2012; hereafter B12) 
performed a similar analysis and found instead that the four- and five-planet cases show about the same success 
rate in matching constraints. Here we discuss the possible causes of this disagreement.
 
B12 used an $N$-body integrator with forces that mimic the effects of gas to generate the initial resonant chains 
of planets. This method is similar to that described for our Phase-1 integrations in Sect. 2, and should 
result in similar initial orbits for Phase 2. The resonant chains explicitly discussed in B12 included 
(3:2,2:1,5:4,5:4) and (3:2,3:2,2:1,4:3). We discussed this type of relaxed resonant chains in Sect. 5.2.2. 
The initial properties of the planetesimal disk in B12 were also similar to the ones used here. We therefore 
believe that the initial conditions are not the main cause of the disagreement. Instead, we explain below that 
different conclusions were reached in B12 probably because B12 gave a different emphasis to different constraints 
(see Sect. 3 for our constraints).

On one hand, B12 found, correcting the results of Batygin \& Brown (2010), that $\sim10$\% of
simulations with four initial planets matched constraints. Their constraints, however, were somewhat different 
and less restrictive than the ones we used here. First of all, B12 did not apply any upper limit on the excitation of  
planetary orbits, while we required in the criterion B that $e<0.11$ and $i<2^\circ$ (relative to the invariant plane).
It is therefore possible that in at least some of the simulations of B12, which gave reasonable $e_{55}$,
some planetary orbits had $e>0.11$ and/or $i>2^\circ$. 

B12 did not consider the constraint from the terrestrial planets (our criterion D). It is therefore
possible that in some of the simulations that were found to be successful in B12, the evolution of the $g_5$ mode
was smooth, which would lead to the secular resonances with the terrestrial planets. Note that some of these models 
could still be valid, but the fraction should be small, probably $<$10\% (Brasser et al. 2009).

While the two issues pointed out above can resolve some of the main differences between B12 and our results,
we are still puzzled, given our clearly negative results for the four-planet case, that B12 were able to 
find positive results for the four-planet case even with very massive disks ($M_{\rm disk} > 50$ M$_{\rm Earth}$).
We tried to repeat the simulations reported in B12 and found that massive disks efficiently damp $e_{55}$ and lead 
to problems with residual migration of Jupiter and Saturn. 

On the other hand, B12 considered constraints from the Kuiper belt that we ignored here. As B12 
pointed out, only about half of their five-planet simulations were compatible with the low eccentricities and 
low inclinations in the cold classical Kuiper belt,\footnote{The cold classical Kuiper Belt is a population of 
trans-Neptunian bodies dynamically defined as having orbits with semimajor axis $a = 42$-48 AU, perihelion distances 
that are large enough to avoid close encounters to Neptune, and low inclinations ($i \lesssim 5^\circ$).}
while most of their four-planet cases were fine. We may have been 
therefore overly optimistic, by a factor of $\sim$2, in favoring the five-planet case over the four-planet case.
The constraints on the planetary instability from the small body reservoirs in the Solar system (Kuiper belt 
objects, asteroids, Trojans and satellites) are clearly very important. We will consider these constraints in 
the follow-up work.  

B12 and this work agree on one central issue. As B12 was not concerned with the delay between the protoplanetary
nebula dispersal and LHB, they placed the planetesimal disk's inner edge just beyond the outer ice giant's orbit. 
This triggered, almost immediately, Neptune's fast migration through the 
disk and the instability mode akin to that we discussed for our most successful runs in Sect. 5. B12 therefore
incidentally explored the setup that we favored here based on a broader sampling of the initial parameters.
It remains to be understood whether this setup can lead to the late instability without overly restrictive 
assumptions on the structure of the planetesimal disk. 

It would be useful in this context, for example, if the LHB started $\sim4.2$ Gyr ago as suggested by Bottke et al. 
(2012), because a shorter time delay since the protoplanetary nebula dispersal ($\simeq$350 Myr) 
would help to relax constraints on $\Delta$.  Alternatively, the planetesimal disk could have been stirred by 
large bodies that formed in it, and 
spread over time, so that the effective $\Delta$ decreased. This could lead to the late instability
even if the initial $\Delta$ was large. Investigations of these issues are left for future work. 
\section{Conclusions}
Recent studies suggest that Jupiter and Saturn formed and migrated in the protoplanetary gas disk to reach a mutual resonance, 
most likely the 3:2 resonance, where the Saturn's orbital period was 3/2 longer than that of Jupiter. After the gas 
disk's dispersal, the orbits of Jupiter and Saturn must have evolved in some way to eventually arrive to the 
current orbits with the orbital period ratio of 2.49. This can most easily be achieved, considering constraints from
the terrestrial planets and $e_{55}$, if Jupiter and Saturn scattered off of Uranus or Neptune, or a planet with mass 
similar to that of Uranus or Neptune. 

We performed $N$-body integrations of the scattering phase between the Solar System's giant planets, including cases 
where one or two extra ice giants were assumed to have formed in the outer Solar System and ejected into 
interstellar space during instability. We found that the initially compact resonant configurations and low masses of 
the planetesimal disk 
($M_{\rm disk}<50$~M$_{\rm Earth}$) typically lead to violent instabilities and planet ejection. On the other hand, 
the initial states with orbits that are more radially spread (e.g., Jupiter and Saturn in the 2:1 resonance) and larger
$M_{\rm disk}$ result in smooth migration of the planetary orbits that leads to incorrectly low $e_{55}$ and excitation 
of the terrestrial planet orbits. Finding the sweet spot between these two extremes is difficult. 

Some of the statistically best results were obtained when assuming that the Solar System initially had five giant planets 
and one ice giant, with the mass comparable to that of Uranus and Neptune, was ejected to interstellar space by Jupiter. 
The best results were obtained when the fifth planet was assumed to have the mass similar to Uranus/Neptune, was placed on 
an orbit just exterior to Saturn's (3:2 and 4:3 resonances work best), and the orbits of Uranus and Neptune migrated
into the planetesimal disk before the onset of planetary scattering. This mode of instability is favored for several 
reasons, as described below.

As planetesimals are scattered by Uranus and Neptune and evolve into the Jupiter/Saturn region, Jupiter, Sa\-turn and the 
fifth planet undergo divergent migration. This triggers an instability during which the fifth planet suffers close 
encounters with all planets and is eventually ejected from the Solar System by Jupiter. Uranus and Neptune generally 
survive the scattering phase, because their orbits migrated outward during the previous stage and opened a protective
gap between them and the gas giants. This mode of instability produces just the right kind of Jupiter's semimajor axis
evolution -known as jumping Jupiter- that is required from the terrestrial planet constraint. 

Moreover, $e_{55}$, excited by the fifth planet ejection, is not damped to incorrectly low values by secular friction
from the planetesimal disk, because the planetesimal disk had been disrupted by Uranus and Neptune before the excitation event. 
The low mass of the planetesimal disk at the time of planet scattering also leads to only a brief migration phase of 
Jupiter and Saturn after the scattering phase, and prevents $P_{\rm Sat}/P_{\rm Jup}$ from evolving beyond its current 
value. The excessive residual migration of Jupiter and Saturn was a problem in most other cases investigated here.

The mode of instability with early migration of Uranus and Neptune is problematic, however, because the inner edge of the 
planetesimal disk may need to be fine-tuned to generate the delay between the formation of the Solar System and the LHB. 
In addition, the range of possible outcomes is rather broad, indicating that the present Solar System is neither typical 
nor an expected result, and occurs, in best cases, with only a $\simeq$5\% probability (as defined by simultaneous matching 
of all four criteria defined in Sect. 3). In $\simeq$95\% of cases, the simulations ended up failing at least one of our
constraints. This may seem unsatisfactory, but given the issues discussed in Sect. 4, the fact that our criteria are 
relatively strict, and because the instability-free model does not work at all,\footnote{The instability-free model
with four planets can be easily tuned to satisfy B, but not C and D, and constraints from the small body 
populations (e.g., Walsh \& Morbidelli 2011, Dawson \& Murray-Clay 2012).} these findings should be seen in a positive 
light. An important follow-up of this work will be to consider additional constraints from the small body populations (e.g., 
the dynamical structure of the Kuiper belt), and see whether the successful simulations identified here will also match those 
constraints.  

The case with six giant planets is also interesting in that the instability occurs in two steps, corresponding to
the ejection of the two planets. The best results were obtained in this case when the two ejected planets were given similar 
masses (about half the mass of Neptune) and were placed between the orbits of Saturn and the inner surviving ice 
giant. As expected, the six-planet case leads to a larger variety of results than the five-planet case. The 
probability of ending the six-planet simulation with the present properties of the solar system is therefore 
lower than in the five-planet case. Still, our six-planet results are fundamentally better than those obtained 
in the four-planet case, where the differences were systematic (e.g., $e_{\rm 55}$ never large enough), and the 
success rate was below the resolution limit of our study. 

\acknowledgments
This work was supported by the NASA's National Lunar Science Institute and Outer Planets Research programs.
Alessandro Morbidelli thanks Germany's Helmholtz Alliance for providing support through its “Planetary 
Evolution and Life” program. David Nesvorny thanks the Observatoire de la C\^ote d'Azur for hospitality 
during his sabbatical in Nice. We also thank an anonymous reviewer for helpful suggestions.

\clearpage
\begin{table}[t]
\begin{center}
\begin{tabular}{rrrr}
\hline
        & $\bar{a}$ (AU) & $\bar{e}$     & $\bar{i}$ ($^\circ$) \\
Jupiter & 5.203  & 0.046 & 0.37 \\
Saturn  & 9.555  & 0.054 & 0.90 \\
Uranus  & 19.22  & 0.044 & 1.02 \\
Neptune & 30.11  & 0.010 & 0.67 \\
\hline
\end{tabular}
\end{center}
\caption{Mean orbital elements of planets. The mean values reported here were obtained by numerically
integrating the orbits of Jupiter, Saturn, Uranus and Neptune for 10 Myr, and computing the average 
of orbital elements over this interval. The mean inclinations are given with respect to the invariant plane.}
\end{table}

\begin{table}[t]
\begin{center}
\begin{tabular}{rrr}
\hline
$j$  & $g_j$ (arcsec yr$^{-1}$)  & $s_j$ (arcsec yr$^{-1}$) \\
5 & 4.24  &  -     \\
6 & 28.22 & -26.34 \\
7 & 3.08  & -2.99  \\
8 & 0.67  & -0.69  \\
\hline
\end{tabular}
\end{center}
\caption{Secular frequencies of giant planets in the Solar System. The frequencies were obtained by numerically
integrating the orbits for 10 Myr, and Fourier analyzing the results. The $s_5$ frequency vanishes when 
the inclinations are referred to the invariant plane.}
\end{table}

\begin{table}[t]
\begin{center}
\begin{tabular}{rrrrr}
\hline
        & 5      & 6      & 7      & 8        \\
Jupiter & {\bf 0.044}  & 0.015  & 0.002 & -        \\
Saturn  & 0.033  & {\bf 0.048}  & 0.002 & -        \\
Uranus  & 0.038  & 0.002 & {\bf 0.029}  & 0.002   \\
Neptune & 0.002  & -      & 0.004  & {\bf 0.009}    \\
\hline
\end{tabular}
\end{center}
\caption{Secular amplitudes $e_{jk}$, where $j$ denotes individual planets (5 to 8 from Jupiter to Neptune).
The proper modes of each planet's orbit are denoted in bold. The amplitudes were obtained by 
numerically integrating the orbits for 10 Myr, and Fourier analyzing the results. Values lower than 
0.001 are not reported.}
\end{table}

\begin{table}[t]
\begin{center}
\begin{tabular}{rrrr}
\hline
        & 6     & 7      & 8        \\
Jupiter & 0.36  & 0.06  & 0.07        \\
Saturn  & {\bf 0.90}  & 0.05  & 0.06        \\
Uranus  & 0.04 & {\bf 1.02}   & 0.06   \\
Neptune & - & 0.12   & {\bf 0.66}    \\
\hline
\end{tabular}
\end{center}
\caption{Secular amplitudes $i_{jk}$. The proper modes of each planet's orbit are denoted in bold. 
The amplitudes are reported in degrees. They were obtained by numerically integrating the orbits for 
10 Myr, and Fourier analyzing the results. Values lower than 0.01$^\circ$ are not reported.
The $s_5$ frequency vanishes in the invariant plane and gives no contribution here.}
\end{table}

\clearpage

\begin{table}[t]
\begin{center}
%\resizebox{!}{12cm} {
\begin{tabular}{rrrrrrrrr}
\hline 
$M_{\rm disk}$     & $\Delta$ & $r_{\rm out}$ & ${\cal B}(j)$ & $N_{\rm sim}$ &  A & B & C & D \\
(M$_{\rm Earth}$)  & (AU)     & (AU)         &               &  & \% & \% & \%  & \% \\
\multicolumn{9}{c}{(3:2,3:2,4:3), $a_4=11.6$ AU}\\      % r32c1 
35   & 0.5 & 30 & 0 & 30  & 13 & 0  & 0    & 0    \\   
35   & 1.0 & 30 & 1 & 30 & 13 & 0  & 0    & 0    \\
50   & 0.5 & 30 & 0 & 30 & 37 & 10 & 0    & 0    \\  % 7,8,22 - Uranus close
50   & 1.0 & 30 & 1 & 100 & 27 & 11 & 1    & 0   \\ 
50   & 3.5 & 30 & 1 & 30 & 13 & 3  & 0    & 0   \\  % 3
50   & 5.0 & 30 & 1 & 30 & 10 & 3  & 0    & 0     \\  % 22
75   & 1.0 & 26 & 1 & 30 & 67 & 40 & 0    & 0    \\  % (run to 30 only), 1,3,5,7,10,13,14,17,18,23,25,26
100  & 1.0 & 25 & 1 & 30 & 80 & 27 & 0    & 0    \\  % 3,6,7,9,11,14,18,27
\multicolumn{9}{c}{(3:2,3:2,3:2), $a_4=12.3$ AU} \\  % r32c3
35   & 1.0 & 30 & 1 & 30 & 40 & 0  & 0    & 0    \\ 
50   & 1.0 & 30 & 1 & 100 & 39 & 4  & 0    & 0    \\  % 28 
\multicolumn{9}{c}{(3:2,4:3,3:2), $a_4=11.9$ AU} \\  % r32c4 (Batygin's res.) 
35     & 1.0 & 30 & 1 & 30 & 7  & 3  & 3  & 0    \\ % 14c <- nice secular system but planets too close 
50     & 1.0 & 30 & 1 & 30 & 10 & 3  & 0  & 0     \\ % 12 (most systems lose a planet)
\multicolumn{9}{c}{(3:2,2:1,2:1), $a_4=18.9$ AU} \\  % r32c5a (pl.in.36, low eccs) <- smooth migration into the disk
35   & 1.0 & 30 & 1 & 100 & 100 & 88 & 0   & 0   \\ % high % for A and B, final systems too cold, S and U too close
50   & 1.0 & 30 & 1 & 100 & 100 & 89 & 0   & 0   \\ % e_J's better only because of smooth crossing of 2:1 between J and S  
\multicolumn{9}{c}{(3:2,2:1,3:2), $a_4=18.9$ AU} \\  % r32c5b (pl.in.104, low eccs)
35   & 1.0 & 30 & 1 & 30 &  0  &  0 &  0  &  0    \\ % 
50   & 1.0 & 30 & 1 & 30 &  0  &  0 &  0  &  0    \\ % 
\multicolumn{9}{c}{(2:1,3:2,3:2), $a_4=14.8$ AU} \\  % r21c1 
35   & 1.0 & 30 & 1 & 100 & 100 & 33 & 11  & 0 \\  % 3,5c,8 (no. 5 is not bad) 
50   & 1.0 & 30 & 1 & 100 & 93  & 87 & 0   & 0  \\  % 1,2,3,4,5,7,8,9,etc <- all bad
\multicolumn{9}{c}{(2:1,3:2,4:3), $a_4=13.7$ AU} \\ % r21c2 
35   & 1.0 & 30 & 1 & 100 & 63 & 13  & 0   & 0  \\  % 10,15
50   & 1.0 & 30 & 1 & 100 & 100 & 50 & 0   & 0  \\  % 2,3,5,7,10,12
\hline
\end{tabular}
%}
\end{center}
\caption{The results of selected four-planet models. The columns are: the (1) mass of the planetesimal disk 
($M_{\rm disk}$), (2)  $\Delta$ defined as $r_{\rm in}-a_n$, where $r_{\rm in}$ is the initial radial 
distance of the inner edge of the planetesimal disk and $a_n$ is the semimajor 
axis of the outer ice giant, (3) radial distance of the outer edge of 
the planetesimal disk ($r_{\rm out}$), (4) ${\cal B}(j)$ specifying the 
instability trigger (see Sect. 2 for a definition), (5) number of simulations performed for each case 
($N_{\rm job}$), and (6-9) success rate for our four criteria defined in Sect. 3.}
\end{table}

\clearpage
\begin{table}[t]
\begin{center}
%\resizebox{!}{24cm} {
\begin{tabular}{rrrrrrrrr}
\hline 
$M_{\rm disk}$     & $\Delta$ & $r_{\rm out}$ & ${\cal B}(j)$ & $N_{\rm sim}$ & A & B & C & D \\
(M$_{\rm Earth}$)  & (AU)     & (AU)         &               &  & \% & \% & \%  & \%      \\
\multicolumn{9}{c}{(3:2,3:2,4:3,5:4), $a_5=13.9$ AU}\\   % r32c1
35   & 1.0 & 30 & 1 & 30  & 13  & 3  & 0    & 3    \\ % 11,12,15d systematic
50   & 1.0 & 30 & 1 & 30  & 37  & 23 & 0    & 0  \\ % <eJ>=0.025, 2,4,7,8,11,16,22 (this one is pretty good but no success on C or D)
50   & 3.5 & 30 & 1 & 30  & 23  & 10 & 0    & 3    \\ % <eJ>=0.024, 13,15d,22 (too many planets lost) 
\multicolumn{9}{c}{(3:2,3:2,3:2,3:2), $a_5=16.1$ AU}  \\ % r32c2 
% This case behaves like Batygin's simulations, instability is generally late with helps e_55
% I could try excited disks for these cases as well
35   & 1.5 & 30 & 0 & 30 & 23 & 3  & 0    & 3   \\ % 3d (very nice jump)
35   & 1.5 & 30 & 1 & 100 & 33 & 16 & 4    & 8    \\ % <eJ>=0.023, 1(d),3,13cd,19cd,21d,30,38,45d <- best!!! 
50   & 1.5 & 30 & 1 & 100 & 30 & 17 & 0    & 3     \\ % 2,7,11(d),28d,29 (Jupiter's eccentricity too low, residual migration)
\multicolumn{9}{c}{(3:2,3:2,4:3,4:3), $a_5=14.5$ AU}  \\ % r32c4a job no. 113451
50   & 1.0 & 30 & 1 & 30 & 47 & 23  & 3     & 3    \\ % <eJ>=0.31, 1,4,5,23cd,26,28,29 (this one has problems with residual migration)
\multicolumn{9}{c}{(3:2,3:2,2:1,3:2), $a_5=22.2$ AU}  \\ % r32c5b, job no. 3. low e's
20     & 1.0 & 30 & 0 & 30 & 33  & 13 &  7 & 3     \\ % 1,5,9c!,28cd! (good but Saturn a bit too close)
20     & 1.0 & 30 & 1 & 30 & 30  & 10 &  7 & 7    \\ % 13cd!,18d,20c (location of Neptune is much better!) <- one of the best 
%20     & 1.0 & 30 & 0 & 43  & 17 &  7 & 3      \\ % 5,6c,14,18,27c!(d) (r32c5b_new2, 0.15, 1000tps) - Saturn a bit close
%20     & 1.0 & 30 & 1 & 23  & 10 &  0 & 3     \\ % 4,9,14d! (r32c5b_new2, 0.15, 1000tps)
35     & 1.0 & 30 & 1 & 30 & 33  & 17 &  0 & 10   \\ % 7d,13d,14d,23,26 (good but excessive damping)
\multicolumn{9}{c}{(3:2,3:2,2:1,2:1), $a_5=24.5$ AU}  \\ % r32c5c, job no. 172, high e's (here the instability happens earlier)
35   & 1.0 & 30 & 0 & 30 & 43  & 17  & 7   & 7       \\ % 2c(d),8,12,24c(d),27d  (disk from 25.5 to 26.5 AU) - Neptune too far
35   & 1.0 & 30 & 1 & 30 & 23  & 13  & 3   & 3       \\ % 2,20,23,24 (disk from 25.5 to 26.5 AU) - Neptune too far
35   & 2.0 & 30 & 1 & 30 & 30  & 23  & 3   & 3        \\ % 6,7,8,18,19cd!,23,30  (disk from 26.5 to 27.5 AU) Neptune too far
35   & 3.0 & 30 & 1 & 100 & 44  & 19  & 3   & 3       \\ % 1,5,10d,15,20        (disk from 27.5 to 28.5 AU) Neptune too far
\multicolumn{9}{c}{(2:1,3:2,3:2,3:2), $a_5=19.3$ AU} \\ % r21c1 (Jup's initial e=0.05) 
20   & 1.0 & 30 & 1 & 30 & 53 & 36 & 3   & 20  \\ % <eJ>=0.020, 6d,8,16d,17d!,21d!,22d,24,25,27d!,29,30 <-best! 
35   & 1.0 & 30 & 1 & 30 & 53 & 43 & 0   & 17   \\ % 2,7,9d,12,14,17d,18d,19,20,25d,26d,28,29  
\multicolumn{9}{c}{(2:1,4:3,3:2,3:2), $a_5=17.9$ AU} \\ % r21c3a (looks like Uranus needs to start near 16.8AU, at least for 20 M_Earth)
20   & 1.0 & 30 & 1 & 30 & 67 & 42 & 0 & 17    \\ % <ej>=0.013,  1d,4d,7,10,12 (U & N do not migrate far enough)
20   & 3.5 & 30 & 1 & 30 & 75 & 25 & 0 & 20   \\ % 5,7d (8 cases only)
\hline
\end{tabular}
%}
\end{center}
\caption{The results of selected five-planet models. See the caption of Table 5 for a definition 
of different parameters shown here.}
\end{table}

\begin{table}[t]
\begin{center}
%\resizebox{!}{12cm} {
\begin{tabular}{rrrrrrrrr}
\hline 
$M_{\rm disk}$     & $\Delta$ & $r_{\rm out}$ & ${\cal B}(j)$ & $N_{\rm sim}$ &  A & B & C & D\\
(M$_{\rm Earth}$)  & (AU)     & (AU)         &               & & \% & \% & \% & \%  \\
\multicolumn{9}{c}{(3:2,3:2,3:2,4:3,3:2), $a_6=20.4$ AU}\\      % r32c1a  r32c1a 
20   & 1.0 & 30 & 1 & 30  & 23  & 7  & 3    & 7   \\ % <eJ>=0.036, 11cd,23d <- nice success for C and D 
35   & 1.0 & 30 & 1 & 30 & 40  & 0  & 0    & 0  \\ % <eJ>=0.044, U & N end up too far  
\multicolumn{9}{c}{(3:2,4:3,3:2,3:2,3:2), $a_6=20.6$ AU}\\      % r32c1b 
20   & 1.0 & 30 & 1 & 100 & 30  & 10  & 3    & 3  \\ % <eJ>=0.04, [1c,8,36,42c(d),47cd,52,91d], p6r32c1   <- best
35   & 1.0 & 30 & 1 & 30 & 42  & 8   & 0    & 3  \\ % <eJ>=0.033, 23d,24
\multicolumn{9}{c}{(2:1,3:2,4:3,3:2,3:2), $a_6=24.2$ AU}\\      % r21c2a (10), very high eccentricity, low masses 
10   & 1.0 & 30 & 1 & 30 & 69  & 25 & 6    & 6  \\ % <eJ>=0.037, 8,10,13d,16c 
20   & 1.0 & 30 & 1 & 30 & 44  & 28 & 3    & 20  \\ % <eJ>=0.024, 1,6d,11d,15d,17d,22,28cd,29d (28 good to 50 My,  great case!)
\multicolumn{9}{c}{(2:1,3:2,4:3,3:2,3:2), $a_6=24.9$ AU}\\      % r21c2b (1328), low masses
20   & 1.0 & 30 & 1 & 30 & 31  & 12 & 3   & 7   \\ % <eJ>=0.038, 15,16cd,17d
\hline
\end{tabular}
%}
\end{center}
\caption{The results of selected six-planet models. 
See the caption of Table 5 for a definition of different parameters shown here. }
\end{table}

\clearpage

\clearpage 

\clearpage

\clearpage

\clearpage 

\clearpage 

\clearpage

\clearpage

\clearpage

\clearpage 

\clearpage 

\clearpage

\clearpage

\clearpage

\clearpage 

\clearpage

\clearpage 

\clearpage


\begin{thebibliography}

\bibitem[Agnor \& Lin(2012)]{2012ApJ...745..143A} Agnor, C.~B., \& Lin, D.~N.~C.\ 2012, \apj, 745, 143 

\bibitem[Batygin 
\& Brown(2010)]{2010ApJ...716.1323B} Batygin, K., \& Brown, M.~E.\ 2010, \apj, 716, 1323 

\bibitem[Batygin et al.(2011)]{2011ApJ...738...13B} Batygin, K., Brown, 
M.~E., \& Fraser, W.~C.\ 2011, \apj, 738, 13 

\bibitem[Batygin et al.(2012)]{2012ApJ...744L...3B} Batygin, K., Brown, 
M.~E., \& Betts, H.\ 2012, \apjl, 744, L3 

\bibitem[Binney \& Tremaine(1987)]{1987gady.book.....B} Binney, J., \& Tremaine, S.\ 1987, Princeton, 
NJ, Princeton University Press.  

\bibitem[Bottke et al.(2012)]{2012Natur.485...78B} Bottke, W.~F., 
Vokrouhlick{\'y}, D., Minton, D., et al.\ 2012, \nat, 485, 78 

\bibitem{bou} Bouvier, A., Blichert-Toft, J., Moynier, F., Vervoort, J. D., Albarede, F. 2007. 
Geochim. Cosmochim. Acta, 71, 1583

\bibitem{bur} Burkhardt, C., Kleine, T., Bourdon, B., Palme, H., Zipfel, J., et al. 2008. 
Geochim. Cosmochim. Acta, 72, 6177

\bibitem[Brasser et 
al.(2009)]{2009A&A...507.1053B} Brasser, R., Morbidelli, A., Gomes, R., Tsiganis, K., \& 
Levison, H.~F.\ 2009, \aap, 507, 1053

\bibitem[Chapman et al.(2007)]{2007Icar..189..233C} Chapman, C.~R., Cohen, 
B.~A., \& Grinspoon, D.~H.\ 2007, Icarus, 189, 233 

\bibitem[Cuk(2007)]{2007DPS....39.6005C} Cuk, M.\ 2007, Bulletin of the 
American Astronomical Society, 38, 537 

\bibitem[Dawson 
\& Murray-Clay(2012)]{2012arXiv1202.6060D} Dawson, R.~I., \& Murray-Clay, R.\ 2012, arXiv:1202.6060 

\bibitem[Duncan et al.(1998)]{1998AJ....116.2067D} Duncan, M.~J., Levison, 
H.~F., \& Lee, M.~H.\ 1998, \aj, 116, 2067 

\bibitem[Goldreich 
\& Sari(2003)]{2003ApJ...585.1024G} Goldreich, P., \& Sari, R.\ 2003, \apj, 585, 1024 

\bibitem[Gomes et al.(2004)]{2004Icar..170..492G} Gomes, R.~S., Morbidelli, 
A., \& Levison, H.~F.\ 2004, Icarus, 170, 492 

\bibitem[Gomes et al.(2005)]{2005Natur.435..466G} Gomes, R., Levison, 
H.~F., Tsiganis, K., \& Morbidelli, A.\ 2005, \nat, 435, 466 

\bibitem{hartman} Haisch, K. E., Jr., Lada, E. A., Lada, C. J. 2001. \apj, 553, L153

\bibitem[Hartmann et al.(2000)]{2000orem.book..493H} Hartmann, W.~K., 
Ryder, G., Dones, L., 
\& Grinspoon, D.\ 2000, Origin of the Earth and Moon, 493 

\bibitem[Jewitt 
\& Haghighipour(2007)]{2007ARA&A..45..261J} Jewitt, D., \& Haghighipour, N.\ 2007, \araa, 45, 261 

\bibitem[Kley(2000)]{2000MNRAS.313L..47K} Kley, W.\ 2000, \mnras, 313, L47 

\bibitem[Lee 
\& Peale(2002)]{2002ApJ...567..596L} Lee, M.~H., \& Peale, S.~J.\ 2002, \apj, 567, 596 

\bibitem[Levison et al.(2001)]{2001Icar..151..286L} Levison, H.~F., Dones, 
L., Chapman, C.~R., et al.\ 2001, Icarus, 151, 286 

\bibitem[Levison et al.(2008)]{2008Icar..196..258L} Levison, H.~F., 
Morbidelli, A., Vanlaerhoven, C., Gomes, R., 
\& Tsiganis, K.\ 2008, Icarus, 196, 258 

\bibitem[Levison et al.(2009)]{2009Natur.460..364L} Levison, H.~F., Bottke, 
W.~F., Gounelle, M., et al.\ 2009, \nat, 460, 364 

\bibitem[Levison et al.(2011)]{2011AJ....142..152L} Levison, H.~F., 
Morbidelli, A., Tsiganis, K., Nesvorn{\'y}, D., 
\& Gomes, R.\ 2011, \aj, 142, 152 

\bibitem[Malhotra(1995)]{1995AJ....110..420M} Malhotra, R.\ 1995, \aj, 110, 
420 

\bibitem[Marcy et al.(2001)]{2001ApJ...556..296M} Marcy, G.~W., Butler, 
R.~P., Fischer, D., et al.\ 2001, \apj, 556, 296 

\bibitem[Masset(2000)]{2000A&AS..141..165M} Masset, F.\ 2000, \aaps, 141, 165 

\bibitem[Masset \& Snellgrove(2001)]{2001MNRAS.320L..55M} Masset, F., \& Snellgrove, M.\ 2001, \mnras, 320, L55 

\bibitem[Minton 
\& Malhotra(2011)]{2011ApJ...732...53M} Minton, D.~A., \& Malhotra, R.\ 2011, \apj, 732, 53 

\bibitem[Morbidelli 
\& Crida(2007)]{2007Icar..191..158M} Morbidelli, A., \& Crida, A.\ 2007, Icarus, 191, 158 

\bibitem[Morbidelli et al.(2005)]{2005Natur.435..462M} Morbidelli, A., 
Levison, H.~F., Tsiganis, K., \& Gomes, R.\ 2005, \nat, 435, 462 

\bibitem[Morbidelli et al.(2007)]{2007AJ....134.1790M} Morbidelli, A., 
Tsiganis, K., Crida, A., Levison, H.~F., \& Gomes, R.\ 2007, \aj, 134, 1790 

\bibitem[Morbidelli et al.(2009)]{2009Icar..202..310M} Morbidelli, A., 
Levison, H.~F., Bottke, W.~F., Dones, L., 
\& Nesvorn{\'y}, D.\ 2009a, Icarus, 202, 310 

\bibitem[Morbidelli et 
al.(2009)]{2009A&A...507.1041M} Morbidelli, A., Brasser, R., Tsiganis, K., Gomes, R., 
\& Levison, H.~F.\ 2009b, \aap, 507, 1041 

\bibitem[Morbidelli et al.(2010)]{2010AJ....140.1391M} Morbidelli, A., 
Brasser, R., Gomes, R., Levison, H.~F., 
\& Tsiganis, K.\ 2010, \aj, 140, 1391 

\bibitem[Nesvorn{\'y} 
\& Vokrouhlick{\'y}(2009)]{2009AJ....137.5003N} Nesvorn{\'y}, D., \& Vokrouhlick{\'y}, D.\ 2009, \aj, 137, 5003 

\bibitem[Nesvorn{\'y} et al.(2007)]{2007AJ....133.1962N} Nesvorn{\'y}, D., 
Vokrouhlick{\'y}, D., \& Morbidelli, A.\ 2007, \aj, 133, 1962 

\bibitem[Nesvorn{\'y}(2011)]{2011ApJ...742L..22N} Nesvorn{\'y}, D.\ 2011, 
\apjl, 742, L22 

\bibitem[Pierens 
\& Raymond(2011)]{2011A&A...533A.131P} Pierens, A., \& Raymond, S.~N.\ 2011, \aap, 533, A131 

\bibitem[Pierens 
\& Nelson(2008)]{2008A&A...482..333P} Pierens, A., \& Nelson, R.~P.\ 2008, \aap, 482, 333 

\bibitem[Rasio 
\& Ford(1996)]{1996Sci...274..954R} Rasio, F.~A., \& Ford, E.~B.\ 1996, Science, 274, 954 

\bibitem{ryder} Ryder, G., 2002, Journal Geophysical Research-Planets, 107, 6

\bibitem[{\v S}idlichovsk{\'y} 
\& Nesvorn{\'y}(1996)]{1996CeMDA..65..137S} {\v S}idlichovsk{\'y}, M., \& Nesvorn{\'y}, D.\ 
1996, Celestial Mechanics and Dynamical Astronomy, 65, 137 

\bibitem[Sumi et al.(2011)]{2011Natur.473..349S} Sumi, T., et al.\ 2011, 
\nat, 473, 349 

\bibitem[Thommes et al.(1999)]{1999Natur.402..635T} Thommes, E.~W., Duncan, 
M.~J., \& Levison, H.~F.\ 1999, \nat, 402, 635 

\bibitem[Thommes et al.(2008)]{2008ApJ...675.1538T} Thommes, E.~W., Bryden, 
G., Wu, Y., \& Rasio, F.~A.\ 2008, \apj, 675, 1538 

\bibitem[Tsiganis et al.(2005)]{2005Natur.435..459T} Tsiganis, K., Gomes, 
R., Morbidelli, A., \& Levison, H.~F.\ 2005, \nat, 435, 459 

\bibitem[Veras 
\& Raymond(2012)]{2012MNRAS.421L.117V} Veras, D., \& Raymond, S.~N.\ 2012, \mnras, 421, L117 

\bibitem[Walsh 
\& Morbidelli(2011)]{2011A&A...526A.126W} Walsh, K.~J., \& Morbidelli, A.\ 2011, \aap, 526, A126 

\bibitem[Walsh et al.(2011)]{2011Natur.475..206W} Walsh, K.~J., Morbidelli, 
A., Raymond, S.~N., O'Brien, D.~P., \& Mandell, A.~M.\ 2011, \nat, 475, 206 

\bibitem[Weidenschilling 
\& Marzari(1996)]{1996Natur.384..619W} Weidenschilling, S.~J., \& Marzari, F.\ 1996, \nat, 384, 619 

\end{thebibliography}
\end{document}